\documentclass[a4paper,11pt]{article}

\usepackage{jcappub} 
\usepackage[T1]{fontenc} 
\usepackage{graphicx}
\usepackage{caption}
\usepackage{subcaption}
\usepackage{float}
\usepackage{slashed}

\title{\boldmath Matter Power Spectrum in Hidden Neutrino Interacting Dark Matter Models: A Closer Look at the Collision Term}

\author[a]{Tobias Binder,}
\author[a]{Laura Covi,}
\author[b]{Ayuki Kamada,}
\author[c,d,e]{Hitoshi Murayama,}
\author[f]{Tomo Takahashi,}
\author[c,g,h]{and Naoki Yoshida}

\affiliation[a]{Institute for Theoretical Physics, Georg-August University G\"ottingen, Friedrich-Hund-Platz 1, G\"ottingen, D-37077 Germany}
\affiliation[b]{Department of Physics and Astronomy, University of California, Riverside, California 92521, USA}
\affiliation[c]{Kavli Institute for the Physics and Mathematics of the Universe (WPI), University of Tokyo Institutes for Advanced Study, University of Tokyo, Kashiwa 277-8583, Japan}
\affiliation[d]{Department of Physics, University of California, Berkeley, Berkeley, California 94720, USA}
\affiliation[e]{Theoretical Physics Group, Lawrence Berkeley National Laboratory, Berkeley, California 94720, USA}
\affiliation[f]{Department of Physics, Saga University, Saga 840-8502, Japan}
\affiliation[g]{Department of Physics, University of Tokyo, Tokyo 113-0033, Japan}
\affiliation[h]{CREST, Japan Science and Technology Agency, 4-1-8 Honcho, Kawaguchi, Saitama, 332-0012, Japan}

\emailAdd{tobias.binder@theorie.physik.uni-goettingen.de}
\emailAdd{laura.covi@theorie.physik.uni-goettingen.de}
\emailAdd{ayuki.kamada@ucr.edu}
\emailAdd{hitoshi.murayama@ipmu.jp}
\emailAdd{tomot@cc.saga-u.ac.jp}
\emailAdd{naoki.yoshida@phys.s.u-tokyo.ac.jp}

\abstract{Dark Matter (DM) models providing possible alternative solutions to the small-scale crisis of standard cosmology are nowadays of growing interest. 
We consider DM interacting with light hidden fermions via well motivated fundamental operators showing the resultant matter power spectrum is suppressed on subgalactic scales within a plausible parameter region. 
Our basic description of the evolution of cosmological perturbations relies on a fully consistent first principles derivation of a perturbed Fokker-Planck type equation, generalizing existing literature. 
The cosmological perturbation of the Fokker-Planck equation is presented for the first time in two different gauges, where the results transform into each other according to the rules of gauge transformation. 
Furthermore, our focus lies on a derivation of a broadly applicable and easily computable collision term showing important phenomenological differences to other existing approximations. 
As one of the main results and concerning the small-scale crisis, we show the equal importance of vector and scalar boson mediated interactions between DM and light fermions.
}

\begin{document}
\maketitle

\def\aj{AJ}%
\def\actaa{Acta Astron.}%
\def\araa{ARA\&A}%
\def\apj{ApJ}%
\def\apjl{ApJ}%
\def\apjs{ApJS}%
\def\ao{Appl.~Opt.}%
\def\apss{Ap\&SS}%
\def\aap{A\&A}%
\def\aapr{A\&A~Rev.}%
\def\aaps{A\&AS}%
\def\azh{AZh}%
\def\baas{BAAS}%
\def\bac{Bull. astr. Inst. Czechosl.}%
\def\caa{Chinese Astron. Astrophys.}%
\def\cjaa{Chinese J. Astron. Astrophys.}%
\def\icarus{Icarus}%
\def\jcap{J. Cosmology Astropart. Phys.}%
\def\jrasc{JRASC}%
\def\mnras{MNRAS}%
\def\memras{MmRAS}%
\def\na{New A}%
\def\nar{New A Rev.}%
\def\pasa{PASA}%
\def\pra{Phys.~Rev.~A}%
\def\prb{Phys.~Rev.~B}%
\def\prc{Phys.~Rev.~C}%
\def\prd{Phys.~Rev.~D}%
\def\pre{Phys.~Rev.~E}%
\def\prl{Phys.~Rev.~Lett.}%
\def\pasp{PASP}%
\def\pasj{PASJ}%
\def\qjras{QJRAS}%
\def\rmxaa{Rev. Mexicana Astron. Astrofis.}%
\def\skytel{S\&T}%
\def\solphys{Sol.~Phys.}%
\def\sovast{Soviet~Ast.}%
\def\ssr{Space~Sci.~Rev.}%
\def\zap{ZAp}%
\def\nat{Nature}%
\def\iaucirc{IAU~Circ.}%
\def\aplett{Astrophys.~Lett.}%
\def\apspr{Astrophys.~Space~Phys.~Res.}%
\def\bain{Bull.~Astron.~Inst.~Netherlands}%
\def\fcp{Fund.~Cosmic~Phys.}%
\def\gca{Geochim.~Cosmochim.~Acta}%
\def\grl{Geophys.~Res.~Lett.}%
\def\jcp{J.~Chem.~Phys.}%
\def\jgr{J.~Geophys.~Res.}%
\def\jqsrt{J.~Quant.~Spec.~Radiat.~Transf.}%
\def\memsai{Mem.~Soc.~Astron.~Italiana}%
\def\nphysa{Nucl.~Phys.~A}%
\def\physrep{Phys.~Rep.}%
\def\physscr{Phys.~Scr}%
\def\planss{Planet.~Space~Sci.}%
\def\procspie{Proc.~SPIE}%
          
\flushbottom

\section{Introduction}
\label{seq:introduction}
Precise measurements of cosmic microwave background (CMB) anisotropies have been building strong evidence for the existence of a new form of matter, called dark matter (DM)~\cite{Hinshaw:2012aka, Ade:2013zuv}.
However, its nature has not yet been uncovered and one of the most important subjects both in astrophysics and in particle physics.
Recently vigorous efforts have been devoted to cosmological probes of interaction strengths between DM and other long-lived particles~\cite{Boehm:2000gq, Chen:2002yh, Sigurdson:2004zp, Boehm:2004th, Mangano:2006mp, Serra:2009uu, Wilkinson:2013kia, CyrRacine:2012fz, Cyr-Racine:2013fsa, Dvorkin:2013cea, Wilkinson:2014ksa, Escudero:2015yka, Ali-Haimoud:2015pwa, Lesgourgues:2015wza}.
Interestingly, such probes are not limited to couplings of the DM to standard model (SM) particles ({\it e.g.}, baryons, photons, and neutrinos).
Couplings to hidden particles are equally subject to searches.
In this paper, we restrict our discussion to hidden light particles, which we call neutrinos for simplicity.
However, the formulation developed and given in this paper is applicable to other models with DM couplings to SM particles.

Interacting DM models are not only within the scope of precise measurements of large-scale structure of the Universe.
They also have their motivation in apparent discrepancies between predictions from DM-only $N$-body simulations and observations on subgalactic scales.
Such discrepancies are called the {\it small-scale crisis} collectively: the {\it missing satellite problem}~\cite{Moore:1999nt, Kravtsov:2009gi}; the {\it cusp vs core problem}~\cite{Moore:1999gc, Donato:2009ab, deBlok:2009sp}; the {\it too big to fail} problem~\cite{BoylanKolchin:2011de, BoylanKolchin:2011dk}.
The simulations assume DM consists of particles with negligible thermal velocities and faint interactions, called cold dark matter (CDM). 
The small-scale crisis may imply alternatives to the CDM paradigm, while it has to be clarified by state-of-the-art hydrodynamic simulations what role baryonic processes play in the formation and evolution of subgalactic objects~\cite{Benson:2001at, Okamoto:2009rw}.
One famous alternative is called the warm dark matter (WDM) model, in which sizable thermal velocities of DM particles suppress the formation of subgalactic objects~\cite{Bode:2000gq}.
Interacting DM effectively reduces the abundance of substructures in a galactic halo to a similar degree as some WDM models do~\cite{Boehm:2001hm, vandenAarssen:2012ag, Aarssen:2012fx, Kamada:2013sh, Boehm:2014vja, Buckley:2014hja, Schewtschenko:2014fca, Cyr-Racine:2015ihg, Vogelsberger:2015gpr, Schewtschenko:2015rno}.

Although there is a growing interest in interacting DM models, it is still unclear what the evolution equations of cosmological perturbations are in such models.
This is because it is difficult to handle and simplify collision terms in Boltzmann equations.
Some works start with the relativistic Navier-Stokes equation for the DM imperfect fluid in particle flow manifest (Eckart's) formulation~\cite{Hofmann:2001bi, Green:2005fa}. 
They determine fluid variables with the help of the Chapman-Enskog method to estimate damping scales in matter power spectra in interacting DM models.
Others just add a collision term in the evolution equations of cosmological perturbations by analogy to the well-known Thomson scattering term for baryons and photons~\cite{Loeb:2005pm}.
One plausible way is to reduce the collision term to the Fokker-Planck equation by assuming the momentum transfer in each collision is smaller than the typical DM momentum.
Such formulation is developed for the traditional bino-like DM in \cite{Bertschinger:2006nq}.
However, the overall factor, {\it i.e.}, the reaction rate, in the Fokker-Planck equation is controversial so far.
A systematic expansion of the collision term in terms of momentum transfer leads to a reaction rate proportional to the invariant amplitude at zero momentum transfer $t \to 0$~\cite{Gondolo:2004sc, Bringmann:2006mu, Bringmann:2009vf}.
On the other hand, in \cite{Kasahara:2009th, Gondolo:2012vh}, the reaction rate is given by $t$-averaging like $\int dt (-t) d\sigma/dt$.

The two formulations introduced above result in different phenomenological consequences.
We consider a simple model, where the SM sector is extended by a Dirac DM, a Dirac (hidden) neutrino, and a mediator. 
A similar scenario is investigated in \cite{vandenAarssen:2012ag}.
When the mediator is a scalar, the reaction rate with zero momentum transfer is negligible and does not change the matter power spectra on and above subgalactic scales within a plausible range of model parameters.
A subgalactic damping scale can be achieved by a vector mediator within this formulation~\cite{Aarssen:2012fx}.
On the other hand, both vector and scalar mediators can suppress the resultant matter power spectra with the 
$t$-averaged reaction rate.
We address this point by calculating the resultant matter power spectra in the latter formulation numerically.
To this end, we derive the evolution equations of cosmological perturbations in two gauges: the conformal Newtonian gauge and the synchronous gauge~\cite{Ma:1995ey}.
We provide an explicit form of gauge transformations between them.
We also show a derivation of the $t$-averaged reaction rate.
It may be useful because a similar derivation is given only in an unpublished thesis~\cite{Kasahara:2009th}.
In the recent ETHOS (effective theory of structure formation) papers~\cite{Cyr-Racine:2015ihg, Vogelsberger:2015gpr}, they study the structure formation in interacting DM models based on the former treatment of the collision term, while they also mention the importance of the $t$-averaging in some models.
The ETHOS paper and this paper are complementary to each other in a treatment of the collision term.

The paper is organized as follows. 
In section~\ref{sec:fokkerplanckmain} we start from first principles and give a detailed derivation of the Fokker-Planck equation with the $t$-averaged reaction rate.
Furthermore, the evolution equations of cosmological perturbations in the synchronous gauge are derived for the most general case of an imperfect fluid. 
We show explicitly in appendix~\ref{subsec:perturbedFPeq} that our results transform into the form of the conformal Newtonian gauge according to the rules of gauge transformation.
In section~\ref{sec:nuinteractingdm}, we give an introduction of the neutrino interacting DM model first.
Then, we summarize our analytic results for scalar and vector mediators: the relic density of DM; the $t$-averaged reaction rate; the resultant smallest mass of halos.
In appendix~\ref{ap:thermalhistory}, we present details of our calculations of chemical decoupling and also summarize the results for models with pseudo scalar and pseudo vector mediators. 
Finally, we show the resultant matter power spectra to stress that not only a vector, but also a scalar mediator can lead to a sizable suppression of matter power spectra. 
In appendix~\ref{ap:higherorder}, we discuss the parameter region for the DM where the perfect fluid approximation is valid. 

Throughout this paper, we use the {\it Planck} 2013 cosmological parameters~\cite{Ade:2013zuv}: $\Omega_{\rm m}=0.3175$, $H_{0}=67.11$, $\ln(10^{10}A_{\rm s})=3.098$, and $n_{\rm s}=0.9624$. 
Updating these input parameters to the {\it Planck} 2015 ones would not change our results significantly.

\section{Fokker-Planck Equation}
\label{sec:fokkerplanckmain}
In this section, the perturbed Fokker-Planck equation is derived. 
Our starting point is the classical Boltzmann equation with the collision term. 
We expand it assuming the momentum transfer per collision is smaller than the typical DM momentum. 
Within this approximation the collision term satisfies detailed balance and respects number conservation. 
As a further result of this expansion method, the momentum transfer rate can easily be computed by first taking a $t$-average and secondly a thermal average of the differential scattering cross section. 
As an important result of the formalism used, the $t$-average is a direct consequence of the expansion method. 
Other methods like in \cite{Bringmann:2006mu} expand the scattering amplitude in terms of small momentum transfer and keep only the zero order. 
But this approximation shows a completely different phenomenology for certain DM theories as will be shown as an explicit example in section~\ref{sec:nuinteractingdm}.
As part of this section, we develop evolution equations of linear cosmological perturbations in the synchronous gauge. 
A comparison to previous works is given. 
The results are equivalent to the conformal Newtonian gauge under the gauge transformation law as we show for the first time in the appendix~\ref{subsec:perturbedFPeq}.

\subsection{Collision Term}
\label{subsec:fokkerplanck}
In this section, we derive a Fokker-Planck equation for the DM phase space distribution function $f$, partially inspired by the unpublished thesis~\cite{Kasahara:2009th}. 
Our starting point is a classical Boltzmann equation for the DM,
\begin{eqnarray}
{\big [} P^{\mu}\partial_{x^{\mu}} - \Gamma^{\mu}_{\kappa \lambda}P^{\kappa}P^{\lambda}\partial_{P^{\mu}} {\big ]} f = C[f] \,,
\end{eqnarray}
where $P_{\mu}$ is the conjugate momentum of the spatial coordinate $x^{\mu}$.
When we handle the collision term $C[f]$, it is convenient to take a local inertial frame $X^{\mu}$, where the metric is $\text{diag}(-1,+1,+1,+1)$ and the proper momentum is denoted by $p^{\mu}=(E, \mathbf{p})$.
We normalize the distribution function such that $\sum_{s} \int d^{3}\mathbf{p}/(2\pi)^{3} (p^{\mu}/E) f = n^{\mu}$, where $s$ are spin degrees of freedom and $n^{\mu}$ is the DM number current.
If we assume the DM particles to interact elastically with particles in a thermal bath, {\it i.e.}, DM(1)+TP(2) $\leftrightarrow$ DM(3)+TP(4), the collision term takes the form
\begin{eqnarray}
C[f_{1}] 
=
\frac{1}{2} 
\sum_{s_{3}}\int \frac{d^{3} \mathbf{p}_{3}}{(2\pi)^{3} 2E_{3}}
{\big [}
-
S^{\rm eq}(p_{1}, p_{3})
f_{1} (1 \mp f_{3})
+
S^{\rm eq}(p_{3}, p_{1})
f_{3} (1 \mp f_{1})
{\big ]}\,,
\end{eqnarray}
where
\begin{eqnarray}
S^{\rm eq}(p_{1}, p_{3}) 
&=&
\sum_{s_{2}}\int \frac{d^{3} \mathbf{p}_{2}}{(2\pi)^{3} 2E_{2}}
\sum_{s_{4}}\int \frac{d^{3} \mathbf{p}_{4}}{(2\pi)^{3} 2E_{4}} 
(2\pi)^{4}\delta^{4}(p_{1} + p_{2} - p_{3} - p_{4})
\notag \\
&& \times
\overline{
|{\cal M}(1+2 \to 3+4)|^{2}
}
f^{\rm eq}_{2} (1 \mp f^{\rm eq}_{4})
\,,\\
S^{\rm eq}(p_{3}, p_{1}) 
&=&
\sum_{s_{2}}\int \frac{d^{3} \mathbf{p}_{2}}{(2\pi)^{3} 2E_{2}}
\sum_{s_{4}}\int \frac{d^{3} \mathbf{p}_{4}}{(2\pi)^{3} 2E_{4}} 
(2\pi)^{4}\delta^{4}(p_{3} + p_{4} - p_{1} - p_{2})
\notag \\
&& \times
\overline{
|{\cal M}(3+4 \to 1+2)|^{2}
}
f^{\rm eq}_{4} (1 \mp f^{\rm eq}_{2})
\,.
\end{eqnarray}
Here, $\overline{|{\cal M}|^{2}}$ is the spin-averaged invariant amplitude squared, and $f^{\rm eq}$ is a thermal distribution,
\begin{eqnarray}
f^{\rm eq} = (\exp\{(-p \cdot u - \mu) / T\} \pm 1)^{-1}
\end{eqnarray}
with a temperature $T \simeq T_{0}(\tau) + T_{1}(x)$, a reference four velocity $u^{\mu} \simeq \left( 1, \mathbf{u}(x) \right)$, and a chemical potential $\mu$.

If the elastic scattering is {\it $T$-inversion invariant}, $\overline{|{\cal M}|^{2}}$'s are identical between the forward and backward scatterings,
\begin{eqnarray}
\overline{
|{\cal M}(1+2 \to 3+4)|^{2}
}
=
\overline{
|{\cal M}(3+4 \to 1+2)|^{2}
}
=
\overline{|{\cal M}|^{2}}
\,.
\label{eq:Tinvariance}
\end{eqnarray}
In the presence of four-momentum conservation $\delta^{4}(p_{1} + p_{2} - p_{3} - p_{4})$, thermal distributions satisfy 
\begin{eqnarray} 
f^{\rm eq}_{2}(1 \mp f^{\rm eq}_{4}) 
= 
\exp\{- (p_{1} - p_{3}) \cdot u / T\}
f^{\rm eq}_{4}(1 \mp f^{\rm eq}_{2})
\,.
\label{eq:relationeq}
\end{eqnarray}
From (\ref{eq:Tinvariance}) and (\ref{eq:relationeq}), we obtain the following relation:
\begin{eqnarray}
S^{\rm eq}(p_{1}, p_{3}) 
=
\exp\{- (p_{1} - p_{3}) \cdot u / T\}
S^{\rm eq}(p_{3}, p_{1}).
\label{eq:seqrelation}
\end{eqnarray}
Thus, the collision term is
\begin{eqnarray}
C[f_{1}] 
=
\frac{1}{2} 
\sum_{s_{3}}\int \frac{d^{3} \mathbf{p}_{3}}{(2\pi)^{3} 2E_{3}}
S^{\rm eq}(p_{3}, p_{1})
{\big [}
f_{3} (1 \mp f_{1})
-
\exp\{- (p_{1} - p_{3}) \cdot u / T\}
f_{1} (1 \mp f_{3})
{\big ]}
\,.
\end{eqnarray}
We can easily check that the above expression satisfies the so-called detailed balance, {\it i.e.}, $C[f_{1}] = 0$ if $f_{1} = f_{1}^{\rm eq}$ and $f_{3} = f_{3}^{\rm eq}$, which follows from the $T$-inversion invariance.

We assume that momentum transfer $\mathbf{\tilde{q}} = \mathbf{p}_{3} - \mathbf{p}_{1}$ is smaller than the typical DM momentum $p_{1 i}$ and expand the collision term up to the second order,
\begin{eqnarray}
f_{3}
\simeq
f_{1}
+
\mathbf{\tilde{q}}_{i} \frac{\partial f_{1}}{\partial \mathbf{p}_{1 i}}
+
\frac{1}{2} \mathbf{\tilde{q}}_{i}\mathbf{\tilde{q}}_{j} \frac{\partial^{2} f_{1}}{\partial \mathbf{p}_{1 i} \partial \mathbf{p}_{1 j}} 
\,, \quad
\exp\{- (p_{1} - p_{3}) \cdot u / T\}
=
1 
+ 
A_{i} \mathbf{\tilde{q}}_{i}
+
B_{i j} \mathbf{\tilde{q}}_{i}\mathbf{\tilde{q}}_{j} \,, 
\notag \\
\label{eq:expfactor}
\end{eqnarray}
where
\begin{eqnarray}
A_{i} = -\frac{\mathbf{v}_{1 i}-\mathbf{u}_{i}}{T}
\,, \quad
B_{i j} = \frac{1}{2} \left(  \frac{\partial A_{i}}{\partial \mathbf{p}_{1 j}} + A_{i} A_{j} \right)
\,,
\end{eqnarray}
with the velocity of the particle $\mathbf{v} = \mathbf{p}/E$.
After collecting terms, we obtain
\begin{eqnarray}
{\big [}
f_{3} (1 \mp f_{1})
-
\exp\{ -(p_{1} - p_{3}) / T\}
f_{1} (1 \mp f_{3})
{\big ]}
\simeq
\alpha_{i} \mathbf{\tilde{q}}_{i}
+
\frac{1}{2}
\left( 
\frac{\partial \alpha_{i}}{\partial \mathbf{p}_{1 j}} 
+ \alpha_{i} A_{j}
\right)
\mathbf{\tilde{q}}_{i} \mathbf{\tilde{q}}_{j}
\,,
\end{eqnarray}
where
\begin{eqnarray}
\alpha_{i} 
= 
\frac{\partial f_{1}}{\partial \mathbf{p}_{1 i}} - A_{i} f_{1} (1 \mp f_{1})\,.
\end{eqnarray}
The collision term is
\begin{eqnarray}
C[f_{1}] 
\simeq
\frac{1}{2} 
{\Big \{}
\alpha_{i} \beta_{i} 
+ 
\frac{1}{2}
\left( 
\frac{\partial \alpha_{i}}{\partial \mathbf{p}_{1 j}} 
+ \alpha_{i} A_{j}
\right)
\gamma_{i j}
{\Big \}} \,,
\label{eq:collisionmid}
\end{eqnarray}
where
\begin{eqnarray}
\label{eq:betagamma}
\beta_{i}
=
\sum_{s_{3}}\int \frac{d^{3} \mathbf{p}_{3}}{(2\pi)^{3} 2E_{3}}
S^{\rm eq}(p_{3}, p_{1})
\mathbf{\tilde{q}}_{i}
\,, \quad
\gamma_{i j}
=
\sum_{s_{3}}\int \frac{d^{3} \mathbf{p}_{3}}{(2\pi)^{3} 2E_{3}}
S^{\rm eq}(p_{3}, p_{1})
\mathbf{\tilde{q}}_{i} \mathbf{\tilde{q}}_{j}
\,.
\end{eqnarray}
It should be noted that $\alpha_{i} = 0$ and thus $C[f_{1}] = 0$ if $f_{1}=f^{\rm eq}_{1}$, which implies that the {\it detailed balance} is maintained in the approximation.

We expand $S^{\rm eq} (p_{3}, p_{1})$ in terms of $\mathbf{\tilde{q}}$, noting that $p_{3}^{\mu}=(\sqrt{E_{1}^2 + 2 \mathbf{p}_{1 i} \mathbf{\tilde{q}}_{i} +  \mathbf{\tilde{q}}^{2}}, \mathbf{p}_1 + \mathbf{\tilde{q}})$.
The scalar function $S^{\rm eq}$ depends on $\mathbf{\tilde{q}}$ only through $\mathbf{p}_{1 i} \mathbf{\tilde{q}}_{i}$ and $\mathbf{\tilde{q}}^{2}$. 
Since we keep the terms only up to the second order in terms of $\mathbf{\tilde{q}}$, the expansion in terms of $\mathbf{\tilde{q}}^{2}$ leads to higher order terms in $C[f_{1}]$, which are to be neglected in our treatment. 
Therefore we expand  $S^{\rm eq}$ only in terms of $\mathbf{p}_{1 i} \mathbf{\tilde{q}}_{i}$ as follows
\footnote{
In fact, if we take $\mathbf{\tilde{q}} \to 0$, $S^{\rm eq}$ diverges owing to a delta function of zero $\delta(0) \delta^{3}(\mathbf{p}_{4} - \mathbf{p}_{2})$ in the integrand.
The expansion just in terms of $\mathbf{p}_{1 i} \mathbf{\tilde{q}}_{i}$ also allows us to avoid such a divergence. 
}:
\begin{eqnarray}
S^{\rm eq}(p_{3}, p_{1}) &\simeq& S^{\rm eq}_{0}(p_{1}, \mathbf{\tilde{q}}^{2}) + S^{\rm eq}_{1}(p_{1}, \mathbf{\tilde{q}}^{2}) \mathbf{p}_{1 i} \mathbf{\tilde{q}}_{i}
\,,  \label{eq:seqapprox1} \\
S^{\rm eq}(p_{1}, p_{3}) &\simeq& S^{\rm eq}_{0}(p_{1}, \mathbf{\tilde{q}}^{2}) 
+ \frac{\partial S^{\rm eq}_{0}(p_{1}, \mathbf{\tilde{q}}^{2})}{\partial \mathbf{p}_{1 i}} \mathbf{\tilde{q}}_{i} 
- S^{\rm eq}_{1}(p_{1}, \mathbf{\tilde{q}}^{2}) \mathbf{p}_{1 i} \mathbf{\tilde{q}}_{i} 
\,,  \label{eq:seqapprox2}
\end{eqnarray}
where $S^{\rm eq}_{0}(p_{1}, \mathbf{\tilde{q}}^{2})$ and $S^{\rm eq}_{1}(p_{1}, \mathbf{\tilde{q}}^{2}) $ are the expansion coefficients defined by  \eqref{eq:seqapprox1}.
 
Substituting (\ref{eq:seqapprox1}) and (\ref{eq:seqapprox2}) into (\ref{eq:betagamma}) and using (\ref{eq:seqrelation}) and (\ref{eq:expfactor}), we obtain
\begin{eqnarray}
\beta_{i} 
&=& 
\sum_{s_{3}}\int \frac{d^{3} \mathbf{p}_{3}}{(2\pi)^{3} 2E_{3}}
\frac{1}{2} 
{\big [}
S^{\rm eq}(p_{3}, p_{1}) 
+ 
\exp\{ (p_{1} - p_{3}) \cdot u / T\}
S^{\rm eq}(p_{1}, p_{3}) 
{\big ]}
\mathbf{\tilde{q}}_{i}
\notag \\
&\simeq&
\sum_{s_{3}}\int \frac{d^{3} \mathbf{p}_{3}}{ 2(2\pi)^{3} }
\left[
\left( - \frac{\mathbf{p}_{1 j} \mathbf{\tilde{q}}_{j}}{E_{1}^{3}} \right)
S^{\rm eq}_{0}(p_{1}, \mathbf{\tilde{q}}^{2})
\mathbf{\tilde{q}}_{i}
+
\frac{1}{2E_{1}} 
(-A_{j} \mathbf{\tilde{q}}_{j})
S^{\rm eq}_{0}(p_{1}, \mathbf{\tilde{q}}^{2})
\mathbf{\tilde{q}}_{i}
+
\frac{1}{2E_{1}}
\frac{\partial S^{\rm eq}_{0}(p_{1}, \mathbf{\tilde{q}}^{2})}{\partial \mathbf{p}_{1 j}}
\mathbf{\tilde{q}}_{i} \mathbf{\tilde{q}}_{j}
\right]
\notag \\
&\simeq&
\frac{E_{1}}{2}
\frac{\partial}{\partial \mathbf{p}_{1 j}}
\left( \frac{1}{E_{1}}
{\tilde \gamma}_{i j}
\right)
-
\frac{1}{2}
A_{j}
{\tilde \gamma}_{i j}\, ,
\label{eq:betaij} 
\end{eqnarray}
where
\begin{eqnarray}
{\tilde \gamma}_{i j}
=
\frac{1}{3}
\delta_{i j}
\sum_{s_{3}}\int \frac{d^{3} \mathbf{p}_{3}}{(2\pi)^{3} 2E_{1}}
S^{\rm eq}_{0}(p_{1}, \mathbf{\tilde{q}}^{2})
\mathbf{\tilde{q}}^{2} \,.
\label{eq:gammaij}
\end{eqnarray}
In the second equality of \eqref{eq:betaij}, we have dropped the term proportional to $S^{\rm eq}_{0}(p_{1}, \mathbf{\tilde{q}}^{2}) \mathbf{\tilde{q}}_{i}$, since it vanishes after integrated in terms of $d^{3} \mathbf{p}_{3}=d^{3} \mathbf{\tilde{q}}$.
In addition, in the last equality, we have replaced  $\mathbf{\tilde{q}}_{i} \mathbf{\tilde{q}}_{j}$ with $(1/3)\delta_{ij}  \mathbf{\tilde{q}}^{2}$ using rotational invariance. 
This is valid under the presence of the integral in terms of $d^{3} \mathbf{p}_{3}=d^{3} \mathbf{\tilde{q}}$.
It should be noted that, $\gamma_{ij} = \tilde{\gamma}_{ij}$ holds up to the second order in $ \mathbf{\tilde{q}}$.

In practice it is more convenient when evaluating (\ref{eq:gammaij}) to replace the perturbative quantities $S^{\rm eq}_{0}(\mathbf{p}_{1}, \mathbf{\tilde{q}}^{2})$ and $-\mathbf{\tilde{q}}^{2}$ with their non-perturbative counterparts $S^{\rm eq}(p_{3}, p_{1})$ and $t / (1-\mathbf{v}_{1 i}\mathbf{v}_{1 i }/3) = -(p_{3}-p_{1})^2 / (1-\mathbf{v}_{1 i}\mathbf{v}_{1 i }/3)$, respectively.
These resulting coefficients only differ through higher order terms and amount to an alternate perturbative expansion.
Substituting (\ref{eq:betaij}) and (\ref{eq:gammaij}) into (\ref{eq:collisionmid}), we obtain a Fokker-Planck-type equation 
for $ f_1$  since the collision term becomes, 
\begin{align}
C[f_{1}]
\simeq
E_{1}
\frac{\partial}{\partial \mathbf{p}_{1 i}}
{\Bigg [}
\gamma
\left (
E_{1} T \frac{\partial f_{1}}{\partial \mathbf{p}_{1 i}} + (\mathbf{p}_{1 i} - E_{1}\mathbf{u}_{i}) f_{1} (1 \mp f_{1})
\right )
{\Bigg ]}
\,,
\end{align}
where the momentum transfer rate is
\begin{eqnarray}
\gamma 
=
\frac{1}{6 E_{1} T (1-\mathbf{v}_{1 i}\mathbf{v}_{1 i }/3)}
\sum_{s_{2}}\int \frac{d^{3} \mathbf{p}_{2}}{(2\pi)^{3}}
f^{\rm eq}_{2} 
(1 \mp f^{\rm eq}_{2})
\int^{0}_{-4\mathbf{p}_{\rm cm}^{2}} dt (-t)
\frac{d\sigma}{dt}v
\,.
\end{eqnarray} 
Here, $d\sigma/dt$ is the differential cross section and $v$ is the relative velocity of the initial particles.
The center of mass momentum is evaluated by $4 s \mathbf{p}_{\rm cm}^{2}=\{ s - (m_{1} - m_{2})^{2} \}\{ s - (m_{1} + m_{2})^{2} \}$, where $s = - (p_{1} + p_{2})^{2}$ and $m$ is the mass of the particle.
This equation satisfies two important requirements.
First, it maintains the detailed balance: if $f_{1} = f_{1}^{\rm eq}$, then $C[f_{1}] = 0$.
Second, it conserves the DM number,
\begin{eqnarray}
\partial_{X^{\mu}} n_{1}^{\mu}
=
\sum_{s_{1}}\int \frac{d^{3} \mathbf{p}_{1}}{(2\pi)^{3}}
\frac{C[f_{1}]}{E_{1}}\
=
0
\,.
\end{eqnarray}

If the DM particles decouple from the thermal bath when they are relativistic, 
momentum transfer in each collision is as large as 
the typical momentum of DM, which may spoil our approximation approach, {\it i.e.}, the Fokker-Planck equation.
It may be useful to give the non-relativistic limit.
Then, the Fokker-Planck equation is 
\begin{eqnarray}
C[f_{1}]
=
m_{1}
\frac{\partial}{\partial \mathbf{p}_{1 i}}
{\Bigg [}
\gamma
\left (
m_{1} T \frac{\partial f_{1}}{\partial \mathbf{p}_{1 i}} + (\mathbf{p}_{1 i} - m_{1}\mathbf{u}_{i}) f_{1}
\right )
{\Bigg ]}
\,,
\end{eqnarray}
where the momentum transfer rate is
\begin{eqnarray}
\gamma 
=
\frac{1}{6 m_{1} T}
\sum_{s_{2}} \int \frac{d^{3} \mathbf{p}_{2}}{(2\pi)^{3}}
f^{\rm eq}_{2} 
(1 \mp f^{\rm eq}_{2})
\int^{0}_{-4\mathbf{p}_{2}^{2}} dt (-t)
\frac{d\sigma}{dt}v
\,.
\label{eq:gammanonrel}
\end{eqnarray}
This expression is the same as given in \cite{Kasahara:2009th, Gondolo:2012vh}.
The cross section is essentially independent of $\mathbf{p}_{1}$ since we consider the case that DM is non-relativistic, or in other words, $|\mathbf{p}_{1}|$ is much smaller than $m_{1}$.
We focus on such a case in the following sections.

Before closing this section, let us discuss the relation between \cite{Gondolo:2004sc, Bringmann:2006mu, Bringmann:2009vf} and the present paper.
The main difference is the presence of $t$-averaging in the momentum transfer rate of (\ref{eq:gammanonrel}).
Once we set $t \to 0$ in $d\sigma/dt$, we can evaluate the $t$-integral analytically to reproduce the result in \cite{Gondolo:2004sc, Bringmann:2006mu, Bringmann:2009vf}.
The $t$-averaging originates from the approximation in (\ref{eq:seqapprox1}) and (\ref{eq:seqapprox2}) and resummation of higher order terms after (\ref{eq:gammaij}).
In this respect, our formulation is not a systematic expansion in terms of the momentum transfer like that in \cite{Bringmann:2006mu}.
However, in some cases, the expansion of invariant amplitudes is not a good approximation  since the leading order does not give the dominant contribution.
One such example is the scalar operator of DM-neutrino interaction investigated in the present paper.
There, the leading order is suppressed by a factor of $m_{\nu}^2/(-t)$ when compared to the next to leading order.

\subsection{Perturbation Theory in the Synchronous Gauge}
\label{subsec:synpert}
Now we develop a linear theory in the synchronous gauge:
\begin{eqnarray}
ds^{2}
=
a^{2}
\left[
-d \tau^{2}
+(\delta_{ij}+h_{ij})d\mathbf{x}^{i}d\mathbf{x}^{j}
\right]\,.
\end{eqnarray}
Up to the first order of cosmological perturbations, the Fokker-Planck equation is given by
\begin{eqnarray}
{\dot f}
+\frac{\mathbf{q}_{i}}{m_{\chi}} \frac{\partial}{\partial \mathbf{x}_{i}} f
-\frac{1}{2} {\dot h}_{ij} \mathbf{q}_{i} \frac{\partial}{\partial \mathbf{q}_{j}} f 
=
(\gamma_{0}+\gamma_{1}) a \frac{\partial}{\partial \mathbf{q}_{i}}
\left[
(\mathbf{q}_{i}-a m_{\chi} \mathbf{u}_{i}) f
+a^{2}m_{\chi} (T_{0}+T_{1}) \frac{\partial}{\partial \mathbf{q}_{i}} f
\right] \,, \notag \\
\end{eqnarray}
with $\gamma = \gamma_{0}(\tau) + \gamma_{1} (x)$ and the comoving momentum $q = a p$
\footnote{Hereafter, for notational simplicity, we respectively use $m_{\chi}$ and $p$ for the DM mass and proper momentum instead of $m_{1}$ and $p_{1}$ that are used in the previous subsection.}.
The homogeneous and isotropic part, {\it i.e.}, the leading order is 
\begin{eqnarray}
{\dot f_{0}}
=
\gamma_{0} a \frac{\partial}{\partial \mathbf{q}_{i}}
\left[
\mathbf{q}_{i} f_{0}
+a^{2}m_{\chi}T_{0} \frac{\partial}{\partial \mathbf{q}_{i}} f_{0}
\right]\,.
\end{eqnarray}
A solution,
\begin{eqnarray}
\label{eq:0thsol}
f_{0}
=
\frac{{\bar n}}{g_{\chi}} \left(\frac{2 \pi}{m_{\chi} T_{\chi 0}}\right)^{3/2} \exp \left( -\frac{\mathbf{q}^{2}}{2 a^{2} m_{\chi} T_{\chi 0}}\right)
\,,
\end{eqnarray}
is parametrized by the DM temperature $T_{\chi 0}(\tau)$ and the DM number density per spin degree of freedom ${\bar n}/g_{\chi}$ with $g_{\chi} = 2 s_{\chi} + 1$.
Its evolution is described by
\begin{eqnarray}
\label{eq:tempevo}
\frac{d \ln (a^{2} T_{\chi 0})}{d \tau}
=
2 \gamma_{0} a \left(\frac{T_{0}}{T_{\chi 0}} -1\right)\,.
\end{eqnarray}
The DM temperature is tightly coupled to the temperature of thermal bath $T_{\chi 0}=T_{0}\propto1/a$ before the kinetic decoupling $\gamma/H >1$.
After they decouple, the DM particles start to stream freely and the temperature decreases adiabatically $T_{\chi 0}\propto1/a^2$.

The first order perturbation follows:
\begin{eqnarray}
\label{eq:f1evolSyn}
{\dot f_{1}}
+\frac{\mathbf{q}_{i}}{m_{\chi}} \frac{\partial}{\partial \mathbf{x}_{i}} f_{1}
-\frac{1}{2} {\dot h}_{ij} \mathbf{q}_{j} \frac{\partial}{\partial \mathbf{q}_{i}} f_{0}
=
\gamma_{1} a L_{\rm FP}[f_{0}]
- \gamma_{0} a^{2} m_{\chi} \mathbf{u}_{i}  \frac{\partial}{\partial \mathbf{q}_{i}} f_{0}
+\gamma_{0} a^{3} m_{\chi} T_{1} \frac{\partial^{2}}{\partial \mathbf{q}^2} f_{0}
+\gamma_{0} a L_{\rm FP}[f_{1}] \,. \notag \\
\end{eqnarray}
Here, we define the Fokker-Planck operator by
\begin{eqnarray}
L_{\rm FP}[f]
=
\frac{\partial}{\partial \mathbf{q}_{i}}
\left[
\mathbf{q}_{i} f
+a^{2}m_{\chi}T_{0} \frac{\partial}{\partial \mathbf{q}_{i}} f
\right] \,.
\end{eqnarray}
In the Fourier space $\mathbf{k}_{i}= k \hat{\mathbf{k}}_{i}$, these equations are rewritten as
\begin{eqnarray}
\label{eq:perturbedFPSyn}
{\dot f_{1}}
+\frac{i \mathbf{k}_{i} \mathbf{q}_{i}}{a m_{\chi}} f_{1}
-\gamma_{0} a L_{\rm FP}[f_{1}]
&& =
{\dot \eta} \frac{\mathbf{q}^{2}}{2 a^{2} m_{\chi} T_{\chi 0}} f_{0}
-\frac{{\dot h} + 6 {\dot \eta}}{2 k^{2}} (\mathbf{k}_{i} \mathbf{q}_{i})^{2} \frac{1}{2 a^{2} m_{\chi} T_{\chi 0}} f_{0}
-\frac{i \mathbf{k}_{i} \mathbf{q}_{i}}{a T_{\chi 0}} \gamma_{0} a \frac{\theta_{\rm TP}}{k^{2}} f_{0}
\notag \\
&& \quad \,
+
\left[ \gamma_{1} a \left(\frac{T_{0}}{T_{\chi 0}} -1\right) + \gamma_{0} a \frac{T_{1}}{T_{\chi 0}} \right] 
\left(  \frac{\mathbf{q}^{2}}{2 a^{2} m_{\chi} T_{\chi 0}} - 3 \right) f_{0} \,.
\end{eqnarray}
Hereafter we consider only the scalar perturbations, defining $\theta_{\rm TP}$, $\eta$, and $h$ such that $\theta_{\rm TP}= i \mathbf{k}_{i}  \mathbf{u}_{i}$ and $h_{i j} = \hat{\mathbf{k}}_{i} \hat{\mathbf{k}}_{j} h + \left( \hat{\mathbf{k}}_{i} \hat{\mathbf{k}}_{j} - \frac{1}{3} \delta_{ij}\right) 6 \eta$ (the same notation as in \cite{Ma:1995ey}).

In order to handle the Fokker-Planck operator, we expand $f_{1}$ in terms of eigenfunctions of the Fokker-Planck operator,
\begin{eqnarray}
L_{\rm FP} \phi_{n\, \ell\, m} 
=
-(2n+\ell) \phi_{n\, \ell\, m}\,,
\quad
\phi_{n\, \ell\, m} 
=
e^{-y} S_{n \ell}(y) Y_{\ell\, m}(\hat{\mathbf{q}}) \,,
\end{eqnarray}
with $y=\mathbf{q}^{2} / (2 a^{2} m_{\chi} T_{0})$, $\mathbf{q}_{i} = |\mathbf{q}| \hat{\mathbf{q}}_{i}$, and a dimensionless function $S_{n\, \ell}(y)=y^{\ell/2} L_{n}^{\ell+1/2}(y)$.
Here $Y_{\ell\, m}$ and $L_{n}^{\alpha}$ denote the spherical harmonics and the Laguerre polynomial, respectively.
Noting the rotational symmetry, we can write
\begin{eqnarray}
&&
f_{1}(\mathbf{k}, \mathbf{q}, \tau)
= 
\frac{1}{(2 \pi a^{2} m_{\chi} T_{0})^{3/2}} e^{-y} \sum_{n,\ell=0}^{\infty} (-i)^{\ell} (2\ell+1) S_{n\, \ell}(y) P_{\ell}(\hat{\mathbf{k}}_{i}\hat{\mathbf{q}}_{i}) f_{n \ell}(k,\tau) \,,
\end{eqnarray}
with the Legendre polynomial $P_{\ell}$, and vice versa,
\begin{eqnarray}
\label{eq:inversion}
f_{n \ell} (k, \tau)
= 
i^{\ell} \frac{\sqrt{\pi}}{2} \frac{n!}{\Gamma(n+\ell+3/2)} \int d^3 \mathbf{q} \, S_{n\, \ell} \left( \frac{\mathbf{q}^{2}}{2 a^{2} m_{\chi} T_{0}} \right) P_{\ell}(\hat{\mathbf{k}}_{i}\hat{\mathbf{q}}_{i}) f_{1}(\mathbf{k}, \mathbf{q}, \tau) \,.
\end{eqnarray}
After a lengthy but straightforward calculation, we obtain the Boltzmann hierarchy:
\begin{eqnarray}
\label{eq:BoltzhierSyn}
&&
{\dot f_{n \ell}}
+(2n+\ell)(\gamma_{0} a + R) f_{n \ell}
-2nRf_{n-1 \ell}
\notag \\
&& \quad
+k \sqrt{\frac{2 T_{0}}{m_{\chi}}} \left\{ \frac{\ell+1}{2\ell+1} \left[ \left(n + \ell + \frac{3}{2} \right) f_{n \ell+1} - n f_{n-1 \ell+1} \right] + \frac{\ell}{2\ell+1} (f_{n+1 \ell-1} - f_{n \ell-1})\right\}
\notag \\
&&\,
=
\delta_{\ell0} 
\left\{ 
-\frac{1}{2} A_{n} {\dot h}  + \frac{1}{3} B_{n} {\dot h}
-2 B_{n} \left[ \gamma_{1} a \left(\frac{T_{0}}{T_{\chi 0}} -1\right) + \gamma_{0} a \frac{T_{1}}{T_{\chi 0}} \right]
\right\}
\notag \\
&& \quad
+
\delta_{\ell1} \frac{1}{3} A_{n} k \sqrt{\frac{2 m_{\chi}}{T_{0}}} \gamma_{0} a \frac{\theta_{\rm TP}}{k^2}
+
\delta_{\ell2} \frac{2}{15} \frac{T_{\chi 0}}{T_{0}} A_{n} ({\dot h} + 6 {\dot \eta})\,.
\end{eqnarray}
Here we introduce three new quantities:
\begin{eqnarray}
\label{eq:quantities}
R
=
\frac{d \ln (a T^{1/2}_{0})}{d \tau} 
\,, \quad
A_{n}
=
\left( 1 - \frac{T_{\chi 0}}{T_{0}} \right)^{n} 
\,, \quad
B_{n}
=
n \frac{T_{\chi 0}}{T_{0}}  \left( 1- \frac{T_{\chi 0}}{T_{0}}  \right)^{n-1} \,.
\end{eqnarray}
The first quantity is essentially proportional to the Hubble expansion rate: $R = a H / 2$.
Only a few of the second and third quantities are non-zero before the kinetic decoupling ($T_{\chi 0 } = T_{0}$): 
$A_{0} = 1$ and $B_{1} = 1$, while the others vanish.
Higher orders of the second quantity become non-zero after the kinetic decoupling ($T_{\chi 0 } \ll T_{0}$):
$A_{n} = 1$, while $B_{n} = n T_{\chi 0 } / T_{0}$ and tiny.

Although we need to solve the full Boltzmann hierarchy to obtain a rigorous result, just taking some small moments of $n$ and $\ell$ can give the {\it fluid} approximation (see discussion in subsection~\ref{subsec:matterpower}).
The perturbations $f_{n \ell}$ with small $n$ and $\ell$ can be interpreted as primitive variables of the DM imperfect fluid ({\it i.e.}, mass density $\rho$, bulk velocity potential $\theta$, pressure $P$ and anisotropic inertia $\sigma$):
\begin{eqnarray}
&&
{\bar \rho} (1 + \delta)
=
-T_{\ 0}^{0}
=
a^{-4} \sum_{s_{\chi}} \int \frac{d^{3}\mathbf{q}}{(2\pi)^3} \, m_{\chi} \, f \,,
\\
&&
({\bar \rho} + {\bar P}) \theta
=
i \mathbf{k}_{i} T_{\ 0}^{i}
=
a^{-4} \sum_{s_{\chi}} \int \frac{d^{3}\mathbf{q}}{(2\pi)^3} \, i \mathbf{k}_{i}\mathbf{q}_{i} \, f \,,
\\
&&
{\bar P} + \delta P
=
\frac{1}{3}T_{\ i}^{i}
=
a^{-4} \sum_{s_{\chi}} \int \frac{d^{3}\mathbf{q}}{(2\pi)^3} \, \frac{\mathbf{q}^{2}}{3 m_{\chi}} \, f \,,
\\
&&
({\bar \rho} + {\bar P}) \sigma
=
- \left( \hat{\mathbf{k}}_{i}\hat{\mathbf{k}}_{j} - \frac{1}{3} \delta_{ij} \right) T_{\ j}^{i}
=
-a^{-4} \sum_{s_{\chi}} \int \frac{d^{3}\mathbf{q}}{(2\pi)^3} \, \frac{\mathbf{q}^{2}}{m_{\chi}} \left[ (\hat{\mathbf{k}}_{i} \hat{\mathbf{q}}_{i})^{2} - \frac{1}{3}\right] \, f \,.
\end{eqnarray}
Substituting the exact form of $f = f_{0}(\tau) + f_{1}$, we obtain
\begin{eqnarray}
&&
{\bar \rho} = m_{\chi} {\bar n} \,, \quad
{\bar P} = \frac{T_{\chi 0}}{m_{\chi}} {\bar \rho} \,,
\\
&&
\label{eq:DMvars}
\delta = f_{0 0} \,, \quad
\theta = 3 k \sqrt{\frac{T_{0}}{2 m_{\chi}}} f_{0 1} \,, \quad
\delta P = \frac{T_{0}}{T_{\chi 0}} \bar{P} (f_{0 0} - f_{1 0}) \,, \quad
\sigma = 5 \frac{T_{0}}{m_{\chi}} f_{0 2} \,.
\end{eqnarray}
The dynamics of the DM imperfect fluid is described by the following equations:
\begin{eqnarray}
&&
{\dot \delta}
=
-\theta
-\frac{1}{2} {\dot h} \,,
\\
&&
{\dot \theta}
=
-\frac{{\dot a}}{a} \theta
-k^{2} \sigma
+k^{2} \frac{T_{\chi 0}}{m_{\chi}} \frac{\delta P}{{\bar P}}
+\gamma_{0} a (\theta_{\rm TP} - \theta) \,,
\\
&&
{\dot \sigma}
=
-2 \frac{{\dot a}}{a} \sigma
-k \left( \frac{2 T_{0}}{m_{\chi}} \right)^{3/2} \left( \frac{21}{4} f_{0 3} + f_{1 1} \right)
+\frac{4}{3} \frac{T_{0}}{m_{\chi}} \theta \,
+\frac{2}{3} \frac{T_{0}}{m_{\chi}} ({\dot h} + 6 {\dot \eta}) 
-2 \gamma_{0} a \sigma \,,
\\
&&
{\dot {\delta P}}
=
-5 \frac{{\dot a }}{a} \delta P
-\frac{5}{6} {\bar P} {\dot h}
+\frac{5}{4} k \left( \frac{2 T_{0}}{m_{\chi}} \right)^{3/2} {\bar \rho} f_{1 1}
-\frac{5}{3} \frac{T_{0}}{T_{\chi 0}} {\bar P} \theta
\notag \\
&& \qquad \ \
-2 \gamma_{0} a \delta P
+2 \gamma_{0} a \frac{T_{0}}{T_{\chi 0}} {\bar P} \delta
+2 {\bar P} \left[ \gamma_{1} a \left( \frac{T_{0}}{T_{\chi 0}} - 1 \right) + \gamma_{0} a \frac{T_{1}}{T_{\chi 0}} \right]\,.
\end{eqnarray}
The pressure perturbation $\delta P$ can be decomposed into isentropic $c_{\chi}^{2} \delta$ and entropy $\pi$ perturbations:
\begin{eqnarray}
\frac{\delta P}{\bar \rho} = c_{\chi}^{2} \delta + \pi \,.
\end{eqnarray}
The sound speed squared of the DM fluid is
\begin{eqnarray}
c_{\chi}^{2} = \frac{T_{\chi 0}}{m_{\chi}} \left(1-\frac{1}{3}\frac{d\ln T_{\chi 0}}{d \ln a} \right)  \,.
\end{eqnarray}
The evolution of the DM imperfect fluid can be rewritten as
\begin{eqnarray}
&&
\label{eq:deltaevolSyn}
{\dot \delta}
=
-\theta
-\frac{1}{2} {\dot h} \,,
\\
&&
\label{eq:thetaevolSyn}
{\dot \theta}
=
-\frac{{\dot a}}{a} \theta
-k^{2} \sigma
+k^{2} (c_{\chi}^{2} \delta + \pi)
+\gamma_{0} a (\theta_{\rm TP} - \theta) \,,
\\
&&
\label{eq:sgimaevolSyn}
{\dot \sigma}
=
-2 \frac{{\dot a}}{a} \sigma
-k \left( \frac{2 T_{0}}{m_{\chi}} \right)^{3/2} \left( \frac{21}{4} f_{0 3} + f_{1 1} \right)
+\frac{4}{3} \frac{T_{0}}{m_{\chi}} \theta \,
+\frac{2}{3} \frac{T_{0}}{m_{\chi}} ({\dot h} + 6 {\dot \eta})
-2 \gamma_{0} a \sigma \,,
\\
&&
\label{eq:pievolSyn}
{\dot \pi}
=
-2 \frac{{\dot a }}{a} \pi
+\frac{5}{4} k \left( \frac{2 T_{0}}{m_{\chi}} \right)^{3/2} f_{1 1}
-\frac{1}{a^{2}} \frac{d (a^{2} c_{\chi}^{2})}{d \tau} \delta
-\left (\frac{5}{3}\frac{T_{0}}{m_{\chi}} - c_{\chi}^{2}\right) \theta
-\frac{1}{2} \left (\frac{5}{3}\frac{T_{\chi 0}}{m_{\chi}} - c_{\chi}^{2}\right) {\dot h}
\notag \\
&& \qquad
-2 \gamma_{0} a \left[ \pi - \frac{T_{1}}{m_{\chi}} - \left( \frac{T_{0}}{m_{\chi}} - c_{\chi}^{2} \right) \delta \right]
+2 \gamma_{0} a \left( \frac{T_{0}}{T_{\chi 0}} - 1 \right) \frac{T_{\chi 0}}{m_{\chi}} \frac{\gamma_{1}}{\gamma_{0}}\,.
\end{eqnarray}

\section{Neutrino Interacting Dark Matter}
\label{sec:nuinteractingdm}
The section starts with the introduction of the neutrino interacting DM model via a MeV-scale boson. 
This particle combination leads in a valid parameter region to a possible solution to all three small-scale crisis problems if the mediator is of vector type~\cite{Aarssen:2012fx}. 
We reproduce and confirm these results by using the method that is derived in the previous section to describe the DM kinetic decoupling.
The used method has a different expansion of the collision term when compared to the aforementioned reference and to others like \cite{Bringmann:2006mu}. 

Furthermore, by using this alternative description we explicitly show a suppression of the power spectrum for other types of mediators as well. 
The suppression is sizable enough to reduce the abundance of dwarf galaxies but unexpected from the point of view of the above literature.
In particular, scalar and vector mediators share an analogue phenomenology within our model set-up and the parameter region is relatively similar concerning the minimal size of the first protohalos.  
Approximation methods to follow the evolution of cosmological perturbations are also given. 
Finally, the matter linear power spectrum for scalar and vector interactions are presented, showing a suppression of powers on subgalactic scales.

\subsection{Simplified Neutrino Model}
\label{subsec:simlifiedmodel}
A simplified model extends SM by a DM fermion and additional light fermions (denoted by $\nu$). 
The DM fermion and the additional light fermions are assumed to be of Dirac type, coupled by a MeV-scale boson denoted by $\phi$. 
In particular, this choice allows us to write down the following set of renormalizable dimension four operators without derivatives:
\begin{align}
\mathcal{L}_{\text{S}} &\supset g_{\chi} \bar{\chi}\phi \chi + g_{\nu} \bar{\nu} \phi \nu \,, \label{eq:operatorlist1} \\
\mathcal{L}_{\text{V}} &\supset g_{\chi}  \bar{\chi}\gamma^{\mu} \chi \phi_{\mu} + g_{\nu}  \bar{\nu} \gamma^{\mu} \nu \phi_{\mu} \,,\label{eq:operatorlist3} \\
\mathcal{L}_{\text{PS}} &\supset g_{\chi}  \bar{\chi}\phi \gamma^5 \chi + g_{\nu}  \bar{\nu}\phi \gamma^5  \nu \,, \label{eq:operatorlist2} \\
\mathcal{L}_{\text{PV}} &\supset g_{\chi}  \bar{\chi}\gamma^{\mu}\gamma^5  \chi \phi_{\mu} +g_{\nu}  \bar{\nu} \gamma^{\mu}\gamma^5  \nu \phi_{\mu} \,. \label{eq:operatorlist4}
\end{align}
Here, we assume parity conservation in the interaction Lagrangian and consider each operator type separately. 
There are four parameters: the DM mass $m_{\chi}$, the light mediator mass $m_{\phi}$, the DM-mediator coupling $g_{\chi}$, and the light fermion-mediator coupling $g_{\nu}$.
Specifically, extensions of the simplified model~(\ref{eq:operatorlist3}) into ultraviolet complete models and constraints have already been investigated by many authors in connection with the small-scale crisis (for an exemplary list of references, see \cite{Dasgupta:2013zpn, Bringmann:2013vra, Ko:2014bka, Cherry:2014xra}).

For simplicity and for analogy to previous works we call the light fermions hidden neutrinos. 
In the early universe, the DM and the hidden neutrinos are assumed to be in thermal equilibrium, where a temperature difference when compared to the SM sector hides the additional light fermions. 
Further, the light boson is in thermal equilibrium with the neutrino sector during the DM chemical freeze-out. 
For all operators the parameters chosen are such that the relic density of the DM is dominantly determined through $\chi \bar{\chi} \to \phi \phi$ annihilation and not via direct $s$-channel neutrino production. 
This is because for the vector, scalar, and pseudo scalar interactions, we assume $g_{\nu} \ll g_{\chi}$ (see \cite{Fox:2008kb} for a list of possible natural explanations). 
In this scenario, the DM relic abundance for all operators is independent of the neutrino coupling $g_{\nu}$. 
In appendix~\ref{subsec:relicabundance} we provide for all operators the full calculus of the annihilation cross section and the relic abundance. 
Due to a more complicated but less illuminating phenomenology, we discuss the results for the pseudo scalar and pseudo vector operators in the appendix~\ref{subsec:parityviolatingoperators}.

\subsection{Minimal Halo Mass}
\label{subsec:minimalhalo}
Elastic scattering via a MeV-scale boson keeps the DM for a long time in kinetic equilibrium with the hidden neutrino sector. 
During kinetic equilibrium, the DM density perturbations do not grow but oscillate.
This phenomena is known as acoustic oscillations and has been shown in \cite{Boehm:2000gq, Boehm:2001hm, Chen:2002yh, Sigurdson:2004zp, Boehm:2004th, Mangano:2006mp, Serra:2009uu, vandenAarssen:2012ag, Aarssen:2012fx, CyrRacine:2012fz, Kamada:2013sh, Wilkinson:2013kia, Cyr-Racine:2013fsa, Dvorkin:2013cea, Wilkinson:2014ksa, Boehm:2014vja, Schewtschenko:2014fca, Cyr-Racine:2015ihg, Vogelsberger:2015gpr, Schewtschenko:2015rno, Escudero:2015yka, Ali-Haimoud:2015pwa} to be the dominant damping mechanism of perturbations in the case of a late kinetic decoupling.

In cosmological perturbation theory, the mode that enters the horizon at the kinetic decoupling defines a cutoff in the linear matter power spectrum of density fluctuations. 
Only the DM density modes that enter the horizon thereafter can significantly grow and collapse later into halos. 
Thus, fluctuations on shorter scales are damped. 
The minimal mass of first protohalos can be estimated by the mass inside a sphere with radius of Hubble horizon at the time of the kinetic decoupling:
\begin{align}
M_{\text{cut}} &= \rho_{\rm m} \frac{4 \pi}{3}\left( \frac{1}{H} \right)^3 = 2.2 \times 10^8 r^3 \left( \frac{1 \,\text{keV}}{T_{\nu}^{\text{kd}}} \right)^{3} \, M_{\odot} \,, 
\label{eq:mcutgleichung}
\end{align}
where the matter density $\rho_{\rm m}$ and the Hubble expansion rate $H$ are evaluated at the kinetic decoupling. 
Here, we allow for a different light fermion temperature from the photon temperature to hide the additional neutrinos. 
The ratio between the two temperatures is defined as $r \equiv T_{\nu}^{\text{kd}} / T_{\gamma}^{\text{kd}}$, where the superscript kd means the corresponding value at the DM kinetic decoupling that occurs when the momentum transfer rate $\gamma$ equals to the Hubble rate $H$.

In the following, we derive an approximation method to estimate the kinetic decoupling temperature $T_{\nu}^{\text{kd}}$ in order to calculate the corresponding cutoff mass according to (\ref{eq:mcutgleichung}). 
The general expression for $\gamma$~(\ref{eq:gammanonrel}) is adjusted to describe the scattering of the DM with the light fermions. 
Dividing it by the Hubble expansion rate and by introducing the following dimensionless variables  $x \equiv |\mathbf{p_{\nu}}|/T_{\nu}$, $y \equiv T_{\nu} / m_{\chi}$ and $z \equiv m_{\phi}/m_{\chi}$, one ends up with the following form: 
\begin{align}
\frac{\gamma}{H}=  \left( \frac{T_{\nu}}{T_{\gamma}}\right)^2 \frac{m_{\text{pl}}}{m_{\chi}} \sqrt{\frac{45}{4 \pi^3}} \frac{N_{\nu}}{48 \pi} \frac{1}{\sqrt{g_{\text{eff}}}} y^{-2} \int_0^{\infty} \text{d}x f_{\nu}^{\text{eq}}(x)\left(1- f_{\nu}^{\text{eq}}(x)\right) g(xy,z) \,, 
\label{eq:gammahubbleexact}
\end{align}
where we multiply by the number of light fermion species $N_{\nu}$. 
The phase-space density function $f_{\nu}^{\text{eq}}(x)$ is the usual equilibrium Fermi-Dirac distribution, where we neglected the mass of the light fermions:
\begin{align}
f_{\nu}^{\text{eq}}(x) = \frac{1}{\exp{(|\mathbf{p_{\nu}}|/T_{\nu}}) + 1} \,.
\end{align}
Furthermore, the dimensionless quantity $g(xy,z)$ is defined as the $t$-averaged scattering amplitude squared:
\begin{align}
g(xy,z) \equiv \frac{1}{m_{\chi}^4(4\pi)^2   } \int_{-4\mathbf{p}_{\nu}^2}^0 \text{d}t (-t)\sum_{s_2,s_3,s_4} \overline{|\mathcal{M}|^2} \,,
\end{align}
where in this convention,
\begin{align}
\overline{|\mathcal{M}|^2} \equiv \frac{1}{16} \sum_{s_1,s_2,s_3,s_4}|\mathcal{M}|^2_{s\rightarrow m_{\chi}^2 + 2 m_{\chi} E_{\nu}}
\end{align}
is the invariant amplitude squared that are averaged over initial {\emph and} averaged over final spin states.

Equation (\ref{eq:gammahubbleexact}) is the basic formula for the kinetic decoupling description of the neutrino interacting DM. 
In the following, we derive analytic estimates for the scalar and vector operators, which are valid in a broad range of parameters and derive their corresponding $M_{\text{cut}}$ scaling patterns. 
In the case of the pseudo scalar and pseudo vector operators, this approximation that we call the effective propagator description is only valid in a small parameter space, and thus (\ref{eq:gammahubbleexact}) has to be solved numerically at some point. 
The results are given in appendix~\ref{subsec:parityviolatingoperators}.
The DM-neutrino scattering amplitudes for the scalar and vector operators are given by:
\begin{align}
\text{Vector operator: }\sum_{s_1,s_2,s_3,s_4}|\mathcal{M}|^2 &= g_{\chi}^2 g_{\nu}^2 \frac{8\left(8 E_{\nu}^2 m_{\chi}^2+4 E_{\nu} m_{\chi} t + t (2 m_{\chi}^2+t)\right)}{(t-m_{\phi}^2)^2} \,, \\
\text{Scalar operator: } \sum_{s_1,s_2,s_3,s_4}|\mathcal{M}|^2 &= g_{\chi}^2 g_{\nu}^2 \frac{4 t (t-4 m_{\chi}^2)}{(t-m_{\phi}^2)^2} \label{eq:scalarscattering}\,.
\end{align}
In the parameter region we are interested in, it turns out that the mass of the mediator is much larger than the kinetic decoupling temperature. 
In this case, the Mandelstam $t$ in the boson propagator denominator of the scattering amplitudes can be neglected. 
We call this approximation the effective propagator description. 
The propagator denominator can be simplified in such a way because $ t \in [0,-4\mathbf{p}^2_{\nu}]$ and the neutrino momentum is further limited by the phase space density suppression: $|\mathbf{p}_{\nu}|  \simeq T_{\nu}$. 
So $t$ can be neglected in the denominator of the propagator as long as $T^{\text{kd}}_{\nu} \ll m_{\phi}$, which is the case in the parameter region of the scalar and vector operators.  

Within the effective propagator framework, $g(xy,z)$ is only a polynomial function in its variables and (\ref{eq:gammahubbleexact}) has even an analytic solution. 
To leading order in  $T_{\nu}$, we find for the vector operator
\begin{align}
\frac{\gamma}{H}= 17.2 \times \left(\frac{r}{r_0}\right)^2 \left(\frac{N_{\nu}}{6}  \frac{\alpha_{\chi}}{0.035}\frac{\alpha_{\nu}}{10^{-4}}\right) \left( \frac{m_{\chi}}{1 \,\text{TeV}} \right)^{-1}\left( \frac{m_{\phi}}{1 \,\text{MeV}} \right)^{-4} \left( \frac{T_{\nu}}{1 \,\text{keV}} \right)^{4} \,,
\end{align}
and for the scalar operator
\begin{align}
\frac{\gamma}{H}= 16.7 \times \left(\frac{r}{r_0}\right)^2 \left(\frac{N_{\nu}}{6}  \frac{\alpha_{\chi}}{0.17}\frac{\alpha_{\nu}}{10^{-5}}\right) \left( \frac{m_{\chi}}{1 \,\text{TeV}} \right)^{-1}\left( \frac{m_{\phi}}{1 \,\text{MeV}} \right)^{-4} \left( \frac{T_{\nu}}{1 \,\text{keV}} \right)^{4} \,,
\end{align}
with $\alpha_{\chi/\nu} = g_{\chi/\nu}^{2}/(4\pi)$.

To estimate the kinetic decoupling temperature, we set $\gamma/H=1$ in the last two equations, which are solved for $T^{\text{kd}}_{\nu}$
\footnote{This defines our kinetic decoupling temperature. 
Another definition of $T_{\text{kd}}$ is used in the literature~\cite{Bringmann:2006mu}, which also has a direct map into the non-linear $M_{\text{cut}}$ given recently in \cite{Cyr-Racine:2015ihg}. 
With our definition, $M_{\text{cut}}$ is smaller by less than a factor of three when compared to aforementioned literature.}. 
The corresponding minimal halo masses that are derived from the kinetic decoupling temperature according to (\ref{eq:mcutgleichung}) is given by: 
\begin{align}
(M_{\text{cut}})_{\text{V}}&=  6.8 \times 10^8 \, M_{\odot} \left(\frac{r}{r_0}\right)^{9/2} \left(\frac{N_{\nu}}{6} \frac{\alpha_{\nu}}{10^{-4}} \frac{\alpha_{\chi}}{0.035} \right)^{3/4} \left( \frac{m_{\chi}}{1 \,\text{TeV}} \right)^{-3/4}\left( \frac{m_{\phi}}{1 \,\text{MeV}} \right)^{-3} \,, 
\label{eq:scalingvector} \\
(M_{\text{cut}})_{\text{S}}&= 6.6 \times 10^8 \, M_{\odot} \left(\frac{r}{r_0}\right)^{9/2} \left(\frac{N_{\nu}}{6} \frac{\alpha_{\nu}}{10^{-5}} \frac{\alpha_{\chi}}{0.17}\right)^{3/4} \left( \frac{m_{\chi}}{1 \,\text{TeV}} \right)^{-3/4}\left( \frac{m_{\phi}}{1 \,\text{MeV}} \right)^{-3} \,,
\label{eq:scalingscalar}
\end{align}
where we normalize $r$ to the SM neutrino temperature ratio: $r_0 = (4/11)^{1/3}$.

To be consistent with constraints on additional radiation components, we use the combined results of Big Bang nucleosynthesis (BBN) and CMB constraints given in \cite{Nollett:2014lwa} to derive an upper bound for our model within the 1$\sigma$ error bar:
\begin{align}
\frac{r}{r_0} < \left( \frac{0.51}{N_{\nu} + \frac{4}{7} g_{\text{pol}} } \right)^{1/4}\,.
\end{align}
Here, we consider the possibility of having a sub-MeV scale mediator contribution to the radiation components at BBN. 
In table~\ref{tab:deltan} we summarize the upper bounds for two extreme scenario: the mediator does not contribute $(g_{\text{pol}}=0)$; the mediator is still relativistic at BBN and contributes via its internal degrees of freedom ($g_{\text{pol}}=\{1,3\}$ for the scalar and massive vector mediators, respectively).

\begin{table}[htb]
\centering
\begin{tabular}{| c | c |}
\hline
$\left(N_{\nu},g_{\text{pol}}\right)$ & $\left( r/r_0 \right)^{9/2} \times\left( N_{\nu}/6 \right)^{3/4} $  \\
\hline
\hline
$\left(2, \{0,1,3 \} \right)$ & $<\left(0.09,0.07,0.05 \right)$ \\
\hline
$\left(6, \{0,1,3 \} \right)$ & $<\left(0.06,0.05,0.05 \right)$ \\
\hline
\end{tabular}
\caption{\label{tab:deltan} Upper bounds on $\left( r/r_0 \right)^{9/2} \times\left( N_{\nu}/6 \right)^{3/4}$ derived from \cite{Nollett:2014lwa} are shown. 
We separate two extreme cases: the mediator is still relativistic at BBN $g_{\text{pol}}=\{1,3\}$; its contribution to the radiation components can be neglected $(g_{\text{pol}}=0)$. 
The factors on the right column reduce the cutoff masses~(\ref{eq:scalingvector}) and (\ref{eq:scalingscalar}) by at least one order of magnitude.}
\end{table}

\begin{figure}[htb]
\centering
 \begin{subfigure}[htb]{0.49\textwidth}
        \includegraphics[scale=0.9]{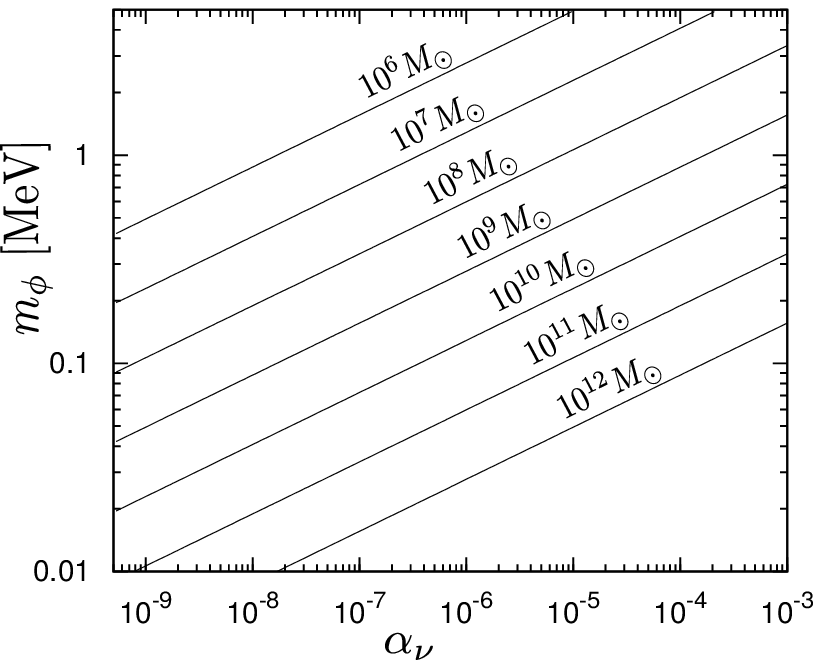}
        \caption{Cutoff mass for the vector mediator}
        \label{fig:vector}
    \end{subfigure}
     \begin{subfigure}[htb]{0.49\textwidth}
        \includegraphics[scale=0.9]{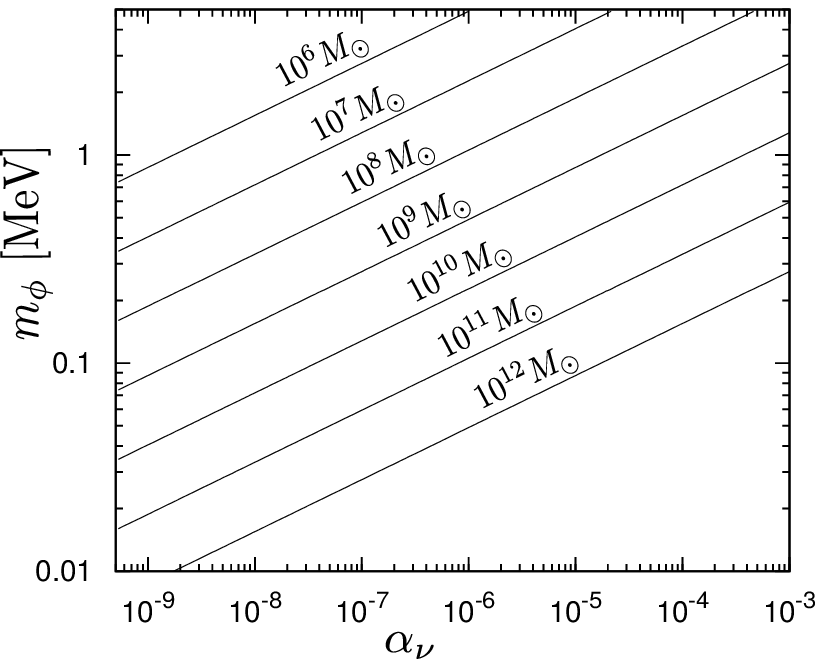}
        \caption{Cutoff mass for the scalar mediator}
        \label{fig:scalar}
    \end{subfigure}
\caption{Contour line of a constant $M_{\text{cut}}$ is shown for the vector (left) and scalar (right) mediators within the effective propagator framework. 
The parameters are chosen according to the normalization values in (\ref{eq:scalingvector}) and (\ref{eq:scalingscalar}). 
In the parameter region shown, the results obtained from the effective propagator description and the exact numerical results obtained by integrating (\ref{eq:gammahubbleexact}) coincide.}
\label{fig:vectorscalarmcuts}
\end{figure}

First of all, these cutoff masses~(\ref{eq:scalingvector}) and (\ref{eq:scalingscalar}) have the same scaling dependence, and thus differ only by a numerical constant and depend mostly on the boson mass. 
Using the relic density constraint on $\alpha_{\chi}$ given by (\ref{eq:relicabundancev}) and (\ref{eq:relicabundances}), we see that $M_{\text{cut}}$ is essentially independent of the DM mass. 
In figure~\ref{fig:vectorscalarmcuts}, contour lines of a constant $M_{\text{cut}}$ are shown for the scalar and vector interactions in the $(m_{\phi},\alpha_{\nu})$-plane. 
In order to account for the missing satellite problem and to be consistent with Ly-$\alpha$ forest bounds, the cutoff mass has to be roughly in between $10^7 \, M_{\odot} \lesssim M_{\text{cut}}\lesssim 5 \times 10^{10} \, M_{\odot}$~\cite{Aarssen:2012fx}. 
We provide the corresponding $M_{\text{cut}}$ contour plots for the pseudo scalar and pseudo vector operators and their discussion in appendix~\ref{subsec:parityviolatingoperators}. 

\subsection{Matter Power Spectrum}
\label{subsec:matterpower}
The minimal halo masses derived in the previous subsection imply that the scalar operator leaves a similar suppression in the resultant matter power spectra to the case of the vector operator. 
In order to see this explicitly, let us consider our model where the DM scatters light fermions via the scalar operator.
The scattering amplitude has a pure $t$-dependence given by (\ref{eq:scalarscattering}). 
In other collision term expansion methods like in \cite{Bringmann:2006mu}, the scattering rate would be declared to be zero at the leading order. 
But as already shown in the previous subsection, we find that DM models with a scalar interaction can also account for the missing satellite problem.

\begin{figure}[htb]
\centering
\includegraphics[scale=1.1]{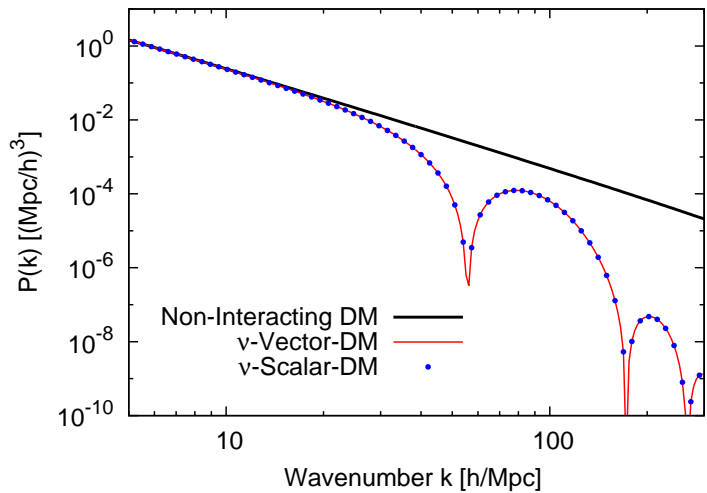}
\caption{This figure shows the linear matter power spectra at present for the standard CDM (black line) and the neutrino interacting DM via the vector (red line) and scalar (blue dots) mediators. 
The wavenumber of $k = 50h\,{\rm /Mpc}$ corresponds to a halo mass of $M = \rho_{\rm m} 4 \pi /3 (\pi/k)^3 \simeq 10^8 \, M_{\odot}$. 
In both the interacting DM cases, we adjust the free neutrino coupling parameters $\alpha_{\nu}$ in (\ref{eq:scalingvector}) and (\ref{eq:scalingscalar}) to give the same cutoff mass ($M_{\text{cut}}= 6.4 \times 10^8 \, M_{\odot}$) and ignored the small effects of the DM-neutrino interactions on the neutrino perturbations ({\it back-reaction}). 
}
\label{fig:matterpowerspectrum}
\end{figure}

To emphasize that scalar interactions are as important as vector interactions regarding the small-scale crisis problems, we adjust the free neutrino coupling parameters $\alpha_{\nu}$ in (\ref{eq:scalingvector}) and (\ref{eq:scalingscalar}) to give the same cutoff mass and show that their linear matter power spectra are close to each other in figure~\ref{fig:matterpowerspectrum}.
Here, we modify the public code \verb|CAMB|~\cite{Lewis:1999bs} suitably to follow the coevolutions of cosmological perturbations of the DM (section~\ref{subsec:synpert}) and the other components ({\it e.g.}, baryon, photon, and gravitational potential).
The small effects of the DM-neutrino interactions on the neutrino perturbations are neglected
and the {\it perfect fluid} approximation (explained below) is used.
Clearly, the shape of the power spectrum shows the characteristic features of the dark acoustic oscillations and the power on small scales is suppressed when compared to the CDM prediction.

Additionally, we check the validity of the perfect fluid assumption by comparing the results to the case of an imperfect fluid.
To obtain a closed set of equations, we need to develop an approximation for $f_{0 3}$ and $f_{1 1}$ (see (\ref{eq:deltaevolSyn})-(\ref{eq:pievolSyn})). 
One way is setting them to be zero, defining the imperfect fluid approximation.
This is valid when $T_{\chi}/m_{\chi} \ll 1$, {\it i.e.}, the free streaming of the DM particles is negligible after they decouple kinetically for $\gamma/H < 1$ (see appendix~\ref{ap:higherorder}).
Actually, we can also take $\sigma=0$ and $\pi=0$ in the same limit, defining the perfect fluid approximation.
Before the kinetic decoupling, all the variables $f_{n \ell}$ but $f_{0 0}$ and $f_{0 1}$ remain zero due to the damping term $\sim \gamma_{0} f_{n \ell}$ in (\ref{eq:BoltzhierSyn}).
The former, corresponding to $\delta$, does not have the damping term.
The latter, corresponding to $\theta$, has the source term $\sim \gamma_{0} (\theta_{\rm TP} - \theta)$.
One non-trivial check is to compare the resultant power spectra in the perfect and imperfect fluid approximations. 

\begin{figure}[htb]
\centering
\includegraphics[scale=1.1]{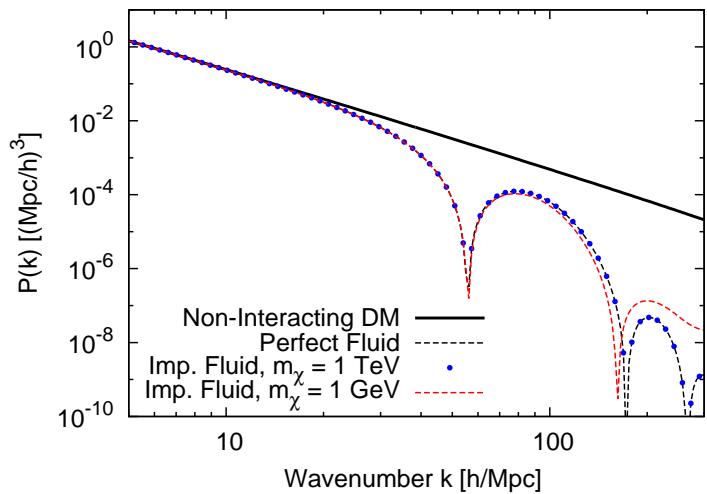}
\caption{This figure compares the linear matter power spectrum in the perfect and imperfect fluid approximations. 
We take the same model with the scalar mediator as in figure~\ref{fig:matterpowerspectrum}. 
We take $m_{\chi}=1\,{\rm TeV}$ in both the perfect and imperfect fluid approximations. 
When the DM mass is lowered to $m_{\chi}=1\,{\rm GeV}$ and $\gamma$ is kept fixed, the resultant matter power spectrum in the imperfect fluid approximation starts to differ from that in the perfect fluid approximation at wavenumbers larger than $k \gtrsim 100 \, h/\text{Mpc}$.}
\label{fig:matterpowerspectrumimperfect}
\end{figure}

When the results from the perfect and imperfect fluid approximations deviate from each other, it does not necessarily mean that the imperfect approximation gives a better description, but it just indicates that the perfect fluid approximation is not valid. 
To check if the imperfect fluid approximation gives a valid description or not, we need to compare the result from the treatment incorporating the full Boltzmann hierarchy, which is beyond the scope of this paper.
Let us stress that the above deviation does not correspond to a limitation of the Fokker-Planck equation, which is valid as long as momentum transfer in each collision is smaller than typical DM momentum.

For a smaller DM mass with $\gamma$ being fixed, we find differences in their power spectrum above a certain critical wavenumber as shown in figure~\ref{fig:matterpowerspectrumimperfect}.
This is because the free streaming is sizable after the kinetic decoupling for the lighter DM.
The results from the perfect fluid approximation are reliable below the critical wavenumber.
On smaller scales, however, we may need to solve the full Boltzmann hierarchy~(\ref{eq:BoltzhierSyn}).
In appendix~\ref{ap:higherorder}, we give a more detailed discussion on the impact of higher order terms in the Boltzmann hierarchy and give a rough estimate of the critical wavenumber, where the results from the perfect and imperfect fluid approximations start to deviate.

\section{Summary and Outlook}
\label{sec:summary}
In summary, we presented a consistent formalism that allows one to start from an underlying DM model and calculate its linear matter power spectrum. 
Regarding the small-scale crisis, the method is broadly applicable to essentially generic radiation interacting DM models that lead to a power spectrum suppression when compared to the standard cosmology on subgalactic scales.

In this paper, we focused on the case where the DM is in kinetic equilibrium with light and hidden fermions for a long time and the decoupling process was investigated for mediators of fundamentally different type. 
The new message is that not only a vector mediator at the MeV scale may solve all three small-scale problems at the same time~\cite{Aarssen:2012fx}, but we find that new classes of interactions may also solve at least the missing satellite problem. 
This result was unexpected from the point of view of previous literature~\cite{Gondolo:2004sc, Bringmann:2006mu, Bringmann:2009vf}, where the leading contribution to the momentum transfer rate is assumed to come from the scattering amplitude evaluated at Mandelstam $t=0$. 
We explicitly derived an expansion method of the collision term where the scattering amplitude is $t$-averaged in the final form of the momentum transfer rate. 
This results in a different phenomenology from that in the previous literature for scattering amplitudes proportional to Mandelstam $t$, {\it e.g.}, in the scalar, pseudo scalar, and pseudo vector interactions between the DM and hidden neutrinos.

With this new insight, the classification of possible DM-radiation interactions, which are suppressing the abundance of dwarf galaxies, has to be revisited. 
During the preparation of this work, we have been informed that Bringmann {\it et al}~\cite{Bringmann:2016ilk} have independently derived similar results concerning the possibility of kinetic decoupling at late times with in new classes of interactions.
As a consequence, our work and studies by the latter authors may extend the list of realistic WIMP-like DM theories accounting for small-scale discrepancies.

As an important subtlety, we also discussed the validity of the perfect fluid approximation for the calculation of the power spectrum. 
We derive the consistent equations needed to be solved for an imperfect fluid treatment and compare the power spectra obtained from the perfect and imperfect fluid approximations. 
As indicated from figure~\ref{fig:matterpowerspectrumimperfect}, the perfect fluid approximation is limited by free streaming effects on the smallest scales. 
This may infer that we need to solve the full Boltzmann hierarchy to have reliable results for some models where the DM mass is small.

Our formalism, as a fundamental building block, in combination with $N$-body simulations would allow one to map DM models into the observational non-linear small-scale structure. 
We plan to combine baryonic feedback and DM induced small-scale suppression to investigate the observational outcome. 
At present or in close future, this kind of sophisticated simulations are expected to shed more light on whether the small-scale crisis will be related to fundamental properties of DM or not. 
Even if the DM-radiation interaction does not resolve the small-scale crisis, our work and others can help to constraint DM models from a new perspective.

\acknowledgments

A.K. would like to thank Paolo Gondolo for kindly providing Junya Kasahara's thesis.
T.B. sincerely thanks Torsten Bringmann for reading our early draft and sharing valuable comments. L.C. would like to thank Giorgio Arcadi for useful discussions. 
T.B. and L.C. acknowledge partial support from the European Union FP7 ITN INVI\-SIBLES (Marie Curie Actions,
PITN-GA-2011-289442). 
The work of T.T. is partially supported by JSPS KAKENHI Grant Number 15K05084 and MEXT KAKENHI Grant Number 15H05888. 
N.Y. acknowledges the financial supports from JST CREST and from JSPS Grant-in-Aid for Scientific Research (25287050, 25610050).

\appendix
\section{Perturbation Theory in the Conformal Newtonian Gauge}
\label{subsec:perturbedFPeq}
In this appendix, we develop a linear theory in the conformal Newtonian gauge and show its equivalence to the synchronous gauge. 
The explicit form of the gauge transformation is presented.
The conformal Newtonian gauge is given by:
\begin{eqnarray}
ds^{2}
=
a^{2}
\left[
-(1+2\Phi)d \tau^{2}
+(1-2\Psi)d\mathbf{x}^{2}
\right]\,.
\end{eqnarray}
Up to the first order of cosmological perturbations, the Fokker-Planck equation in the conformal Newtonian gauge is given by:
\begin{eqnarray}
&&
{\dot f}
+\frac{\mathbf{q}_{i}}{m_{\chi}} \frac{\partial}{\partial \mathbf{x}_{i}} f
+\left({\dot \Psi} \mathbf{q}_{i} - m_{\chi} \frac{\partial}{\partial \mathbf{x}_{i}}\Phi \right) \frac{\partial}{\partial \mathbf{q}_{i}} f 
\notag \\
&&\,
=
(\gamma_{0}+\gamma_{1}) a (1 + \Phi) \frac{\partial}{\partial \mathbf{q}_{i}}
\left[
(\mathbf{q}_{i}-am_{\chi} \mathbf{u}_{i}) f
+a^{2}m_{\chi} (T_{0}+T_{1}) \frac{\partial}{\partial \mathbf{q}_{i}} f
\right]\,.
\end{eqnarray}

Between the conformal Newtonian and synchronous gauges, the collision term differs by a factor of $(1 + \Phi)$\footnote{
The factor $(1 + \Phi)$ is missing in the corresponding equation in \cite{Bertschinger:2006nq}.
}.
This is because in the conformal newtonian gauge, the gravitational potential $\Phi$ put the conformal time back in relative to the local inertial time.
The first order perturbation follows:
\begin{eqnarray}
&&
{\dot f_{1}}
+\frac{\mathbf{q}_{i}}{m_{\chi}} \frac{\partial}{\partial \mathbf{x}_{i}} f_{1}
+\left({\dot \Psi} \mathbf{q}_{i} - m_{\chi} \frac{\partial}{\partial \mathbf{x}_{i}}\Phi \right) \frac{\partial}{\partial \mathbf{q}_{i}} f_{0}
\notag \\
&&\,
=
(\gamma_{1} + \gamma_{0} \Phi) a L_{\rm FP}[f_{0}]
- \gamma_{0} a^{2} m_{\chi} \mathbf{u}_{i}  \frac{\partial}{\partial \mathbf{q}_{i}} f_{0}
+\gamma_{0} a^{3} m_{\chi} T_{1} \frac{\partial^{2}}{\partial \mathbf{q}^{2}} f_{0}
+\gamma_{0} a L_{\rm FP}[f_{1}]\,.
\end{eqnarray}
In the Fourier space, these equations are rewritten as
\begin{eqnarray}
\label{eq:perturbedFPCon}
{\dot f_{1}}
+\frac{i \mathbf{k}_{i} \mathbf{q}_{i}}{a m_{\chi}} f_{1}
-\gamma_{0} a L_{\rm FP}[f_{1}]
&&=
{\dot \Psi} \frac{\mathbf{q}^{2}}{2 a^{2} m_{\chi} T_{\chi 0}} f_{0}
-\frac{i \mathbf{k}_{i} \mathbf{q}_{i}}{a T_{\chi 0}} \left( \Phi + \gamma_{0} a \frac{\theta_{\rm TP}}{k^{2}} \right) f_{0}
\notag \\
&& \quad \,
+
\left[ (\gamma_{1} + \gamma_{0} \Phi) a \left(\frac{T_{0}}{T_{\chi 0}} -1\right) + \gamma_{0} a \frac{T_{1}}{T_{\chi 0}} \right] 
\left(  \frac{\mathbf{q}^{2}}{2 a^{2} m_{\chi} T_{\chi 0}} - 3 \right) f_{0} \,. \notag \\
\end{eqnarray}
We obtain the Boltzmann hierarchy,
\begin{eqnarray}
\label{eq:BoltzhierCon}
&&
{\dot f_{n \ell}}
+(2n+\ell)(\gamma_{0} a + R) f_{n \ell}
-2nRf_{n-1 \ell}
\notag \\
&& \quad
+k \sqrt{\frac{2 T_{0}}{m_{\chi}}} \left\{ \frac{\ell+1}{2\ell+1} \left[ \left(n + \ell + \frac{3}{2} \right) f_{n \ell+1} - n f_{n-1 \ell+1} \right] + \frac{\ell}{2\ell+1} (f_{n+1 \ell-1} - f_{n \ell-1})\right\}
\notag \\
&& \,
=
\delta_{\ell0} 
\left\{ 
3 A_{n} {\dot \Psi}
-2 B_{n} \left[ {\dot \Psi} + (\gamma_{1} + \gamma_{0} \Phi) a \left(\frac{T_{0}}{T_{\chi 0}} -1\right) + \gamma_{0} a \frac{T_{1}}{T_{\chi 0}} \right]
\right\}
\notag \\
&& \quad
+
\delta_{\ell1} \frac{1}{3}  A_{n} k \sqrt{\frac{2 m_{\chi}}{T_{0}}} \left( \Phi + \gamma_{0} a \frac{\theta_{\rm TP}}{k^{2}} \right) \,.
\end{eqnarray}
This description is equivalent to that in the synchronous gauge through the gauge transformation of 
\begin{eqnarray}
&&
f_{n 0}({\rm Syn}) 
= 
f_{n 0}({\rm Con}) + (3 A_{n} - 2 B_{n}) \frac{{\dot a}}{a} \alpha + B_{n} \frac{d \ln (a^{2} T_{\chi 0})}{d \tau} \alpha \,,
\\
&&
f_{n 1}({\rm Syn}) 
= 
f_{n 1}({\rm Con}) - \frac{1}{3} A_{n} k \sqrt{\frac{2 m_{\chi}}{T_{0}}} \alpha \,,
\\
&&
T_{1}({\rm Syn}) 
=
T_{1}({\rm Con})  - {\dot T_{0}} \alpha \,,
\\
&&
\gamma_{1}({\rm Syn}) 
=
\gamma_{1}({\rm Con})  - {\dot \gamma_{0}} \alpha \,,
\end{eqnarray}
with the parameter $\alpha=({\dot h} + 6 {\dot \eta})/(2 k^{2})$.
Here, it should be noted again that the above gauge transformation works only with the time delay of the collision term $(1 + \Phi)$ in the conformal Newtonian gauge.
The dynamics of the DM imperfect fluid is described by the following equations:
\begin{eqnarray}
&&
{\dot \delta}
=
-\theta
+3 {\dot \Psi} \,,
\\
&&
{\dot \theta}
=
-\frac{{\dot a}}{a} \theta
-k^{2} \sigma
+k^{2} \frac{T_{\chi 0}}{m_{\chi}} \frac{\delta P}{{\bar P}}
+k^{2} \Phi
+\gamma_{0} a (\theta_{\rm TP} - \theta) \,,
\\
&&
{\dot \sigma}
=
-2 \frac{{\dot a}}{a} \sigma
-k \left( \frac{2 T_{0}}{m_{\chi}} \right)^{3/2} \left( \frac{21}{4} f_{0 3} + f_{1 1} \right)
+\frac{4}{3} \frac{T_{0}}{m_{\chi}} \theta
-2 \gamma_{0} a \sigma \,,
\\
&&
{\dot {\delta P}}
=
-5 \frac{{\dot a }}{a} \delta P
+5 {\bar P} {\dot \Psi}
+\frac{5}{4} k \left( \frac{2 T_{0}}{m_{\chi}} \right)^{3/2} {\bar \rho} f_{1 1}
-\frac{5}{3} \frac{T_{0}}{T_{\chi 0}} {\bar P} \theta
\notag \\
&& \qquad \ \
-2 \gamma_{0} a \delta P
+2 \gamma_{0} a \frac{T_{0}}{T_{\chi 0}} {\bar P} \delta
+2 {\bar P} \left[ (\gamma_{1} + \gamma_{0} \Phi) a \left( \frac{T_{0}}{T_{\chi 0}} - 1 \right) + \gamma_{0} a \frac{T_{1}}{T_{\chi 0}} \right]\,.
\end{eqnarray}
The evolution of the DM imperfect fluid can be rewritten with isentropic and entropy perturbations:
\begin{eqnarray}
\label{eq:contieq}
&&
{\dot \delta}
=
-\theta
+3 {\dot \Psi} \,,
\\
\label{eq:Eulereq}
&&
{\dot \theta}
=
-\frac{{\dot a}}{a} \theta
-k^{2} \sigma
+k^{2} (c_{\chi}^{2} \delta + \pi)
+k^{2} \Phi
+\gamma_{0} a (\theta_{\rm TP} - \theta) \,,
\\
&&
{\dot \sigma}
=
-2 \frac{{\dot a}}{a} \sigma
-k \left( \frac{2 T_{0}}{m_{\chi}} \right)^{3/2} \left( \frac{21}{4} f_{0 3} + f_{1 1} \right)
+\frac{4}{3} \frac{T_{0}}{m_{\chi}} \theta
-2 \gamma_{0} a \sigma \,,
\\
\label{eq:entroeq}
&&
{\dot \pi}
=
-2 \frac{{\dot a }}{a} \pi
+\frac{5}{4} k \left( \frac{2 T_{0}}{m_{\chi}} \right)^{3/2} f_{1 1}
-\frac{1}{a^{2}} \frac{d (a^{2} c_{\chi}^{2})}{d \tau} \delta
-\left (\frac{5}{3}\frac{T_{0}}{m_{\chi}} - c_{\chi}^{2}\right) \theta
+3\left (\frac{5}{3}\frac{T_{\chi 0}}{m_{\chi}} - c_{\chi}^{2}\right) {\dot \Psi}
\notag \\
&& \qquad
-2 \gamma_{0} a \left[ \pi - \frac{T_{1}}{m_{\chi}} - \left( \frac{T_{0}}{m_{\chi}} - c_{\chi}^{2} \right) \delta \right]
+2 \gamma_{0} a \left( \frac{T_{0}}{T_{\chi 0}} - 1 \right) \frac{T_{\chi 0}}{m_{\chi}} \left( \frac{\gamma_{1}}{\gamma_{0}} + \Phi \right) \,.
\end{eqnarray}

\section{Impact of the Higher Order Terms in the Boltzmann Hierarchy}
\label{ap:higherorder}
In this appendix we take a closer look at the higher order terms in the Boltzmann hierarchy.
As discussed in subsection~\ref{subsec:matterpower}, they represent the free streaming of DM particles and are important for the case of a smaller DM mass.
Once we solve the full Boltzmann hierarchy directly, we can see their effects on resultant matter power spectra quantitatively.
It is, however, challenging and beyond the scope of this paper.
Instead we give an estimate of the critical wavenumber, below which the perfect fluid approximation appears trustworthy.

Before the kinetic decoupling ($\gamma/H \gg 1$), the higher order terms are negligible.
This is because the friction term ($\propto \gamma_{0} a$) in the Boltzmann hierarchy~(\ref{eq:BoltzhierSyn}) leads $f_{n \ell}$ to a rapid damping:
\begin{eqnarray}
{\dot f_{n \ell}}
+(2n+\ell) \gamma_{0} a f_{n \ell}
=
-2 \delta_{n 1} \delta_{\ell 0} 
\gamma_{0} a \frac{T_{1}}{T_{\chi 0}}
+
\delta_{\ell1} \frac{1}{3} A_{n} k \sqrt{\frac{2 m_{\chi}}{T_{0}}} \gamma_{0} a \frac{\theta_{\rm TP}}{k^2} \,.
\end{eqnarray}
Here we have used $T_{\chi 0} = T_{0}$, which results in $B_{1} = 1$ and $B_{n} = 0$ ($n \neq 1$) as discussed below (\ref{eq:quantities}).
Exceptions are $f_{0 0}$, $f_{0 1}$, and $f_{1 0}$ since the first does not have the friction term in its evolution equation and the last two have the source terms (right-handed side) induced by the collision ($\propto \gamma_{0} a$) in their evolution equations.
Through (\ref{eq:DMvars}), $f_{0 0}$, $f_{0 1}$, and $f_{1 0}$ are respectively related with the density perturbation $\delta$, the bulk velocity $\theta$, and the entropy perturbation $\pi$.
From (\ref{eq:pievolSyn}) with a rapid momentum transfer, 
\begin{eqnarray}
{\dot \pi}
=
-2 \gamma_{0} a \left( \pi - \frac{T_{1}}{m_{\chi}} + \frac{T_{0}}{3 m_{\chi}} \delta \right) \,,
\end{eqnarray}
we can see that the entropy perturbation is proportional to the isocurvature perturbation $S_{\rm TP, DM} = \delta(s/n)/({\bar s}/{\bar n}) =  3 T_{1}/T_{0} - \delta$: $\pi = 1/3 (T_{0}/m_{\chi}) S_{\rm TP, DM}$.
As long as DM particles and those in thermal bath are tightly coupled to each other, thereby forming a one fluid, $S_{\rm TP, DM}$ vanishes for adiabatic perturbations.
Thus only $f_{0 0}$ and $f_{0 1}$, or in other words, $\delta$ and $\theta$ are non-zero.
The perfect fluid approximation is valid before the kinetic decoupling.

After the kinetic decoupling ($\gamma/H \ll 1$), higher order terms become sizable.
They, however, do not change the resultant matter power of long wavelength modes as follows.
In this limit, we can neglect the term proportional to $k \sqrt{T_{0}/m_{\chi}}$:
\begin{eqnarray}
{\dot f_{n \ell}}
+(2n+\ell)R f_{n \ell}
-2nRf_{n-1 \ell}
=
-\frac{1}{2} \delta_{\ell0} {\dot h}  
+
\delta_{\ell1} \frac{1}{3} k \sqrt{\frac{2 m_{\chi}}{T_{0}}} \gamma_{0} a \frac{\theta_{\rm TP}}{k^2}
+
\delta_{\ell2} \frac{2}{15} \frac{T_{\chi 0}}{T_{0}} ({\dot h} + 6 {\dot \eta}) \,. \notag \\
\end{eqnarray}
Here we have used $A_{n} = 1$ and $B_{n} \ll 1$ after the kinetic decoupling (see the discussion below (\ref{eq:quantities})).
Noting that $f_{n-1 \ell}$ affects the evolution of $f_{n \ell}$ through the term of $-2nRf_{n-1 \ell}$, we can see that the higher order terms $f_{n 0}$ become of the order of $f_{0 0} = \delta$ within a few Hubble time after the kinetic decoupling.
This, however, does not affect the evolution of $\delta$ and thus does not change the resultant matter power.
This is because $f_{n-1 \ell}$ affects the evolution of $f_{n \ell}$ but not vice versa.

From the above observations, we infer that the impact of higher order terms is suppressed by a factor of $k \sqrt{T_{0}/m_{\chi}} / (aH)$.
Thus we can estimate the critical wavenumber by equating the factor with unity.
This ratio scales in proportion to $a^{1/2}$ ($a^{0}$, or in other words, constant) in the radiation (matter) dominated era, and hence it takes a maximum value of $k/k_{\rm eq} \sqrt{T_{0}(a_{\rm eq})/m_{\chi}}$ with the wavenumber $k_{\rm eq}$ and scale factor $a_{\rm eq}$ at the matter radiation equality.
As a result, we infer that for 
\begin{eqnarray}
k \ll 430\,/{\rm Mpc}\times\,\left(\frac{r}{r_{0}}\right)^{-1/2}\left(\frac{m_{\chi}}{{\rm GeV}}\right)^{1/2} \,,
\end{eqnarray}
the fluid approximation is trustworthy (see discussion below (\ref{eq:mcutgleichung}) for the definition of $r$).
In figure~\ref{fig:matterpowerspectrumimperfect}, the deviation between the results from the perfect and imperfect fluid approximations can be seen above $k\simeq100\,h/$Mpc for $m_{\chi}=1$\,GeV.
This appears compatible with the above estimation.

\section{Thermal History Calculation}
\label{ap:thermalhistory}
In this appendix, the annihilation cross section and relic abundance are presented for all the operators. 
The minimal halo mass and momentum transfer rates of the pseudo scalar and pseudo vector operators are presented as well, showing a different kind of phenomenology when compared to the scalar and vector ones.

\subsection{Relic Abundance}
\label{subsec:relicabundance}
In our simplified model, the DM abundance is dominantly determined via annihilation process into two mediators $\phi$. 
The invariant amplitude for this process is a sum over $t$- and $u$-channel diagrams. 
In order to calculate the DM relic abundance, the cross section times relative velocity $(\sigma v_{\text{rel}})$ is expanded to the leading order in terms of the relative velocity $v_{\text{rel}}$ and mass ratio $z\equiv m_{\phi}/m_{\chi}$.
For each operator, the expanded annihilation cross section is given by:
\begin{align}
(\sigma v_{\text{rel}})_{\text{V}}&= \frac{g_{\chi}^4}{16 \pi m_{\chi}^2} \left(1-\frac{1}{2} z^2 + \mathcal{O}(z^4) \right) + \frac{g_{\chi}^4}{16 \pi m_{\chi}^2} \left(\frac{19}{24} z^2  + \mathcal{O}(z^4)\right)v_{\text{rel}}^2 + \mathcal{O}(v_{\text{rel}}^4) \,, 
\label{mvectorcs} \\
(\sigma v_{\text{rel}})_{\text{S}}&= \frac{3 g_{\chi}^4  }{128 \pi m_{\chi}^2} \left(1 + \frac{11}{18} z^2 + \mathcal{O}(z^4) \right) v_{\text{rel}}^2 + \mathcal{O}(v_{\text{rel}}^4) \,, 
\label{eq:scalarcs} \\
(\sigma v_{\text{rel}})_{\text{PV}}&= \frac{g_{\chi}^4}{16 \pi m_{\chi}^2} \left(1-\frac{1}{2} z^2 + \mathcal{O}(z^4) \right) + \frac{g_{\chi}^4}{12 \pi m_{\chi}^2} \left( z^{-4} + \mathcal{O}(z^{-2})\right) v_{\text{rel}}^2 + \mathcal{O}(v_{\text{rel}}^4) \,, 
\label{eq:pvectorcs} \\
(\sigma v_{\text{rel}})_{\text{PS}}&= \frac{ g_{\chi}^4  }{384 \pi m_{\chi}^2} \left(1 - \frac{1}{2} z^2 + \mathcal{O}(z^4) \right) v_{\text{rel}}^2 + \mathcal{O}(v_{\text{rel}}^4) \,. 
\label{eq:pscalarcs}
\end{align}
The scalar, vector, and pseudo scalar cross sections are consistent with the ones obtained in \cite{Liu:2013vha}. 
In the case of the pseudo vector interaction, we find the leading term to be proportional to $z^{-4}$.
We discuss this subtlety in subsection~\ref{subsec:parityviolatingoperators}.

We estimate the DM freeze-out temperature $x_f$, following basically the method used in \cite{Cannoni:2015wba},  and determine the relic abundance for each operator, given approximately by:
\begin{align}
\text{V: }\Omega_{\chi}h^2 &= \frac{0.12}{2} \left( \frac{\alpha_{\chi}}{0.035} \right)^{-2} \left( \frac{m_{\chi}}{1 \,\text{TeV}} \right)^2 \left( \frac{x_f}{26.1} \right) \,, 
\label{eq:relicabundancev} \\
\text{S: }\Omega_{\chi}h^2 &= \frac{0.12}{2} \left( \frac{\alpha_{\chi}}{0.17} \right)^{-2} \left( \frac{m_{\chi}}{1 \,\text{TeV}} \right)^2 \left( \frac{x_f}{26.8} \right)^2  \,, 
\label{eq:relicabundances} \\
\text{PV: }\Omega_{\chi}h^2 &= \frac{0.12}{2} \left( \frac{r}{r_0} \right) \left( \frac{\alpha_{\chi}}{8.4 \times 10^{-12}} \right)^{-2} \left( \frac{m_{\chi}}{100 \,\text{MeV}} \right)^{2} \left( \frac{z}{10^{-3}} \right)^4 \left( \frac{x_f}{13.4} \frac{r_0}{r} \right)^2 \,, 
\label{eq:relicabundancepv} \\
\text{PS: }\Omega_{\chi}h^2 &= \frac{0.12}{2} \left( \frac{r}{r_0} \right) \left( \frac{\alpha_{\chi}}{4.7 \times 10^{-5}} \right)^{-2} \left( \frac{m_{\chi}}{100 \,\text{MeV}} \right)^2 \left( \frac{x_f}{13.4} \frac{r_0}{r} \right)^2 \,. 
\label{eq:relicabundanceps}
\end{align}
In the case of the scalar and vector operators, we assume the DM, $\phi$, and the light fermions to have the same temperature as the SM particles at the DM freeze-out. 
In the case of the pseudo scalar and pseudo vector operators, we had to lower the DM mass in order to get a cutoff mass around $ \sim 10^8 \,M_{\odot}$ as shown in subsection~\ref{subsec:parityviolatingoperators}. 
The DM freeze-out in this case occurs at a time close to BBN. 
Thus, the temperature of $\phi$ and light fermions has to differ from the SM radiation temperature in order to be hidden and not to be in conflict with observation. 
This subtlety is taken into account in the relic abundance of (\ref{eq:relicabundancepv}) and (\ref{eq:relicabundanceps}). 
Throughout this paper, we ignore the logarithmic dependence of the freeze-out temperature $x_f/r$ on the model parameters and drop the last factor in (\ref{eq:relicabundancev})-(\ref{eq:relicabundanceps}) if the relic density constraint is used to reduce one of the parameters.

Furthermore, we remark that due to the presence of a light mediator and its long range property one has to include the Sommerfeld effect for DM annihilation in principle. 
This may lead to a $\mathcal{O}(1)$ correction of the DM coupling in order to produce the correct relic abundance, but including the effect is beyond the scope of this paper.

\subsection{Minimal Halo Mass of Pseudo Scalar and Pseudo Vector Operators}
\label{subsec:parityviolatingoperators}
In the case of the pseudo scalar and pseudo vector operators, the parameter space of interest spoils partially the effective propagator description, and thus $\gamma/H$ does not have a simple power law dependence on the neutrino temperature like in the scalar and vector cases. 
Nevertheless, we derive analytically the scaling pattern of the cutoff mass from the effective propagator description, and compare it to the cutoff mass derived from the exact numerical evaluation of (\ref{eq:gammahubbleexact}).
The DM-neutrino scattering amplitudes for the pseudo scalar and pseudo vector operators are given by:
\begin{align}
\text{Pseudo scalar operator: } \sum_{s_1,s_2,s_3,s_4}|\mathcal{M}|^2 &= g_{\chi}^2 g_{\nu}^2 \frac{4 t^2 }{(t-m_{\phi}^2)^2} \,, 
\label{eq:pseudoscalarscattering} \\
\text{Pseudo vector operator: } \sum_{s_1,s_2,s_3,s_4}|\mathcal{M}|^2 &= g_{\chi}^2 g_{\nu}^2 \frac{8\left(8 E_{\nu}^2 m_{\chi}^2+4 E_{\nu} m_{\chi} t - t (2 m_{\chi}^2-t)\right)}{(t-m_{\phi}^2)^2} \,. 
\label{eq:pseudovectorscattering}
\end{align}

\subsubsection*{Pseudo Scalar Operator}
\begin{figure}[htb]
\centering
 \begin{subfigure}[htb]{0.49\textwidth}
        \includegraphics[scale=0.88]{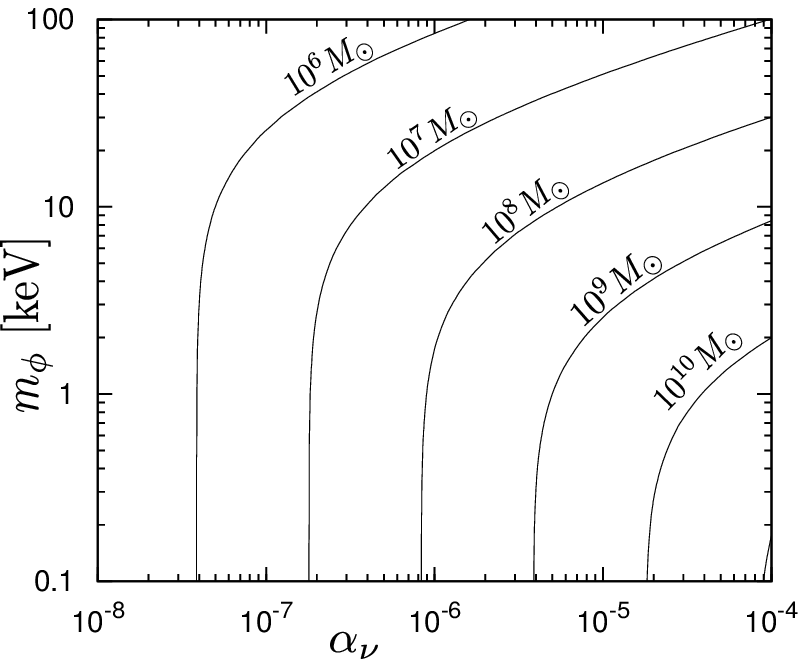}
        \caption{Exact numerical result}
        \label{fig:exactpseudoscalar}
    \end{subfigure}
     \begin{subfigure}[htb]{0.49\textwidth}
        \includegraphics[scale=0.88]{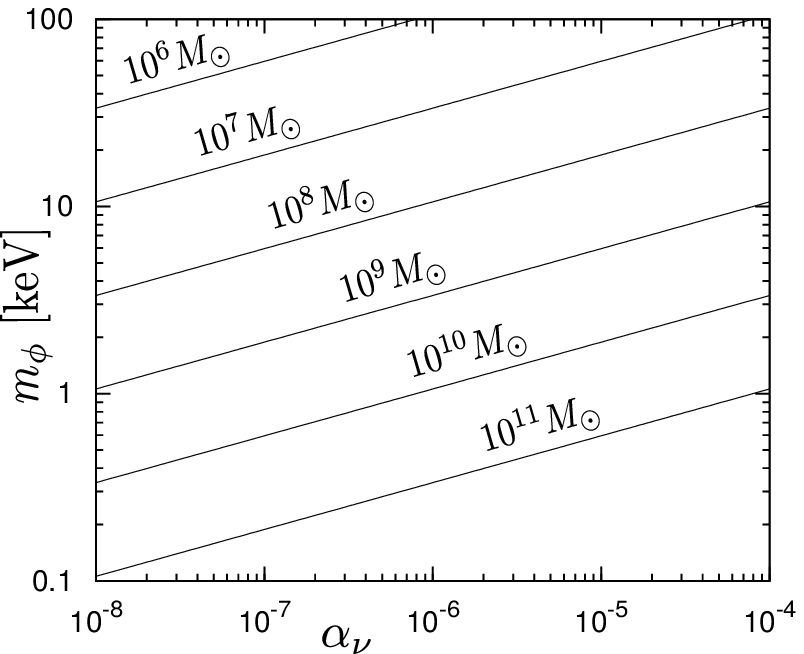}
        \caption{Effective propagator description}
        \label{fig:effectivepseudoscalar}
    \end{subfigure}
\caption{Contour line of a constant $M_{\text{cut}}$ is shown for the pseudo scalar operator within the exact (left) and effective propagator framework (right). 
The DM parameters chosen are $m_{\chi}=100$\,MeV and $\alpha_{\chi}$ satisfying the relic density constraint.
 The effective propagator description is only valid in the upper right quarter of figure~\ref{fig:effectivepseudoscalar}.}
\label{fig:pseudoscalarmcuts}
\end{figure}

The DM-neutrino scattering amplitude~(\ref{eq:pseudoscalarscattering}) via a pseudo scalar mediator has a pure $t^2$ dependence. 
Within the effective propagator framework, $\gamma/H$ depends therefore on a different power of $T_{\nu}$ when compared to the scalar and vector operators:
\begin{align}
\frac{\gamma}{H}= 2.0 \times \left(\frac{r}{r_0}\right)^2  \left(\frac{N_{\nu}}{6} \frac{\alpha_{\chi}}{4.7 \times 10^{-5}}\frac{\alpha_{\nu}}{10^{-6}}\right) \left( \frac{m_{\chi}}{100 \,\text{MeV}} \right)^{-3}\left( \frac{m_{\phi}}{10 \,\text{keV}} \right)^{-4} \left( \frac{T_{\nu}}{1 \text{keV}} \right)^{6} \,.
\end{align}
Inserting the relic density constraint for $\alpha_{\chi}$ given by (\ref{eq:relicabundanceps}), we find the scaling pattern of the cutoff mass:
\begin{align}
(M_{\text{cut}})_{\text{PS}}= 1.1 \times 10^8  \, M_{\odot} \left(\frac{r}{r_0}\right)^{15/4} \left(\frac{N_{\nu}}{6} \frac{\alpha_{\nu}}{10^{-6}}\right)^{1/2} \left( \frac{m_{\chi}}{100 \,\text{MeV}} \right)^{-1}\left( \frac{m_{\phi}}{10 \,\text{keV}} \right)^{-2} \,.
\end{align}
Note that the mass of the mediator is close to the temperature $\sim 1$\,keV for subgalacitc cutoff masses.
This spoils our effective propagator description as can be seen by comparing the exact numerical result with the effective description in figure~\ref{fig:pseudoscalarmcuts}. 

\subsubsection*{Pseudo Vector Operator}
The DM annihilation cross section via a pseudo vector mediator shows a $z^{-4}$ enhancement in (\ref{eq:pvectorcs}). 
At a first look, the limit $z \rightarrow 0$ in the cross section seems to diverge and give rise to unitarity violation~\cite{Kahlhoefer:2015bea}.
By embedding the model into a local $U(1)$ gauge theory where both the mass of the DM and the gauge boson mass arise due to the spontaneous symmetry breaking via an additional scalar field, we show explicitly that this is not the case and the parameter region that we use to produce subgalactic cutoffs is in the perturbative regime.

We denote the additional scalar by $\Phi$ and the local $U(1)$ gauge invariant action reads:
\begin{align}
\mathcal{L}&= i \bar{\chi} \slashed{D}_+\chi +  |D_{\mu,-2}\Phi|^2 - \frac{1}{4} F_{\mu\nu}F^{\mu\nu} - \lambda_Y \left( \bar{\chi}_L \Phi \chi_R + \bar{\chi}_R \Phi^{\star} \chi_L \right) - V(\Phi) \,,
\end{align}
where $\slashed{D}_+=  \slashed{\partial} + i g_{\chi} \slashed{\phi} \gamma^5 $, $D_{\mu,-2}=  \partial_{\mu} - i 2 g_{\chi} \phi_{\mu} $, $V(\Phi)=-\mu^2 \Phi^{\star} \Phi + \frac{\lambda}{2} \left(\Phi^{\star} \Phi\right)^2$ and the fields transform such that 
\begin{align}
\chi \rightarrow e^{i \gamma^5 \alpha(x)}\chi\,, \,\, \phi_{\mu} \rightarrow \phi_{\mu} - \frac{1}{g_{\chi}} \partial_{\mu} \alpha(x)\,,\,\,\Phi \rightarrow e^{-2 i \alpha(x)} \Phi \,.
\end{align}
The vacuum expectation value of the field $\Phi$ in this potential is given by $v\equiv \sqrt{\frac{\mu^2}{\lambda}}$. We expand the scalar field around its minimum $\Phi(x) = v + \frac{1}{\sqrt{2}} \left(h(x) + i \Phi_2(x)\right)$ and get the following relevant quantities after symmetry breaking: $m_{\chi} \equiv \lambda_Y v$, $\frac{m_{\phi}^2}{2} \equiv 4 g_{\chi}^2 v^2$, scalar mass $m_h = \sqrt{2 \lambda v^2} =\sqrt{2} \mu$, Yukawa interaction $- \frac{\lambda_Y}{\sqrt{2}} \bar{\chi} h \chi  =  - \frac{m_{\chi}}{v \sqrt{2}} \bar{\chi} h \chi $, and scalar-gauge boson interaction $ + 4 \sqrt{2} g_{\chi}^2 v \, h \phi_{\mu} \phi^{\mu}$.

The invariant amplitude of DM annihilation into two gauge bosons contains three terms:
\begin{align}
\mathcal{M}= &\epsilon^{\star}_{\mu}(k_2)\epsilon^{\star}_{\nu}(k_1)\bar{v}(p_2) \left(-i g_{\chi} \gamma^{\mu} \gamma^5\right) \frac{i\left(\slashed{p_1} -\slashed{k_1} + m_{\chi}\right)}{\left(p_1-k_1\right)^2 - m_{\chi}^2} \left(-i g_{\chi} \gamma^{\nu} \gamma^5 \right)u(p_1) \notag \\
&+\epsilon^{\star}_{\mu}(k_1)\epsilon^{\star}_{\nu}(k_2)\bar{v}(p_2) \left(-i g_{\chi} \gamma^{\mu} \gamma^5\right) \frac{i\left(\slashed{p_1} -\slashed{k_2}+ m_{\chi}\right)}{\left(p_1-k_2\right)^2 - m_{\chi}^2} \left(-i g_{\chi} \gamma^{\nu} \gamma^5 \right)u(p_1) \notag \\
&+ 2\times \bar{v}(p_2) \left(-i \frac{m_{\chi}}{\sqrt{2}v}\right) u(p_1) \frac{i}{\left(p_1+p_2\right)^2 - m_h^2}\left(+i 4 \sqrt{2} g_{\chi}^2 v \right) \epsilon^{\star}_{\mu}(k_2)\epsilon^{\star,\mu}(k_1) \,,
\label{eq:scalarcontr} 
\end{align}
and the total cross section results in
\begin{align}
(\sigma v_{\text{rel}}) = &\frac{g_{\chi}^4}{16 \pi m_{\chi}^2}\left(1-\frac{1}{2}z^2+\mathcal{O}(z^4)\right) \notag \\
 &+ \frac{g_{\chi}^4}{16 \pi (y^2-4)^2 m_{\chi}^2} \left(\frac{4}{3}(y^4+8)z^{-4}-\frac{16}{3}(2 y^2+1)z^{-2}+\mathcal{O}(z^0)\right)v_{\text{rel}}^2+\mathcal{O}(v_{\text{rel}}^4) \,,
\label{eq:leadingterm1} 
\end{align}
where $z=m_{\phi}/m_{\chi}$ and $y=m_{h}/m_{\chi}$.
Now, the limit of $m_{\phi} \rightarrow 0$ ($z\rightarrow 0$), effectively meaning $g_{\chi} \rightarrow 0$, results in a finite value of the annihilation cross section that is proportional to $\lambda_{Y}^4$.
In the following, we show that all parameters are in the perturbative regime and the scalar contribution~(\ref{eq:scalarcontr}) can be ignored in the low energy expansion, so that (\ref{eq:leadingterm1}) reduces to  (\ref{eq:pvectorcs}).

When we choose $z \sim 10^{-3}$, due to the $z^{-4}$ enhancement in the annihilation cross section~(\ref{eq:pscalarcs}), the DM coupling is forced to be tiny in order to satisfy the relic abundance constraint.
A choice of $m_{\chi}=100$\,MeV leads to $g_{\chi}=1.0 \times 10^{-5}$. 
With these choices, we derive $\lambda_Y = 0.03$, $v=\frac{m_{\chi}}{\lambda_Y}= 3.4 \,\text{GeV}$, and $y=\frac{\sqrt{2 \lambda}}{\lambda_Y} \lesssim 49$. 
If we take $y$ of $\mathcal{O}(10)$, we see directly that the leading term in (\ref{eq:leadingterm1}) is indeed given by
\begin{align}
(\sigma v_{\text{rel}}) \simeq \frac{g_{\chi}^4}{12 \pi m_{\chi}^2} z^{-4} v_{\text{rel}}^2  \,.
\label{eq:leadingterm2}
\end{align}
This result is the same as the leading one in (\ref{eq:pvectorcs}) and the relic abundance given by (\ref{eq:relicabundancepv}), where the scalar contribution has been ignored, holds.

\begin{figure}[htb]
\centering
 \begin{subfigure}[htb]{0.49\textwidth}
        \includegraphics[scale=0.9]{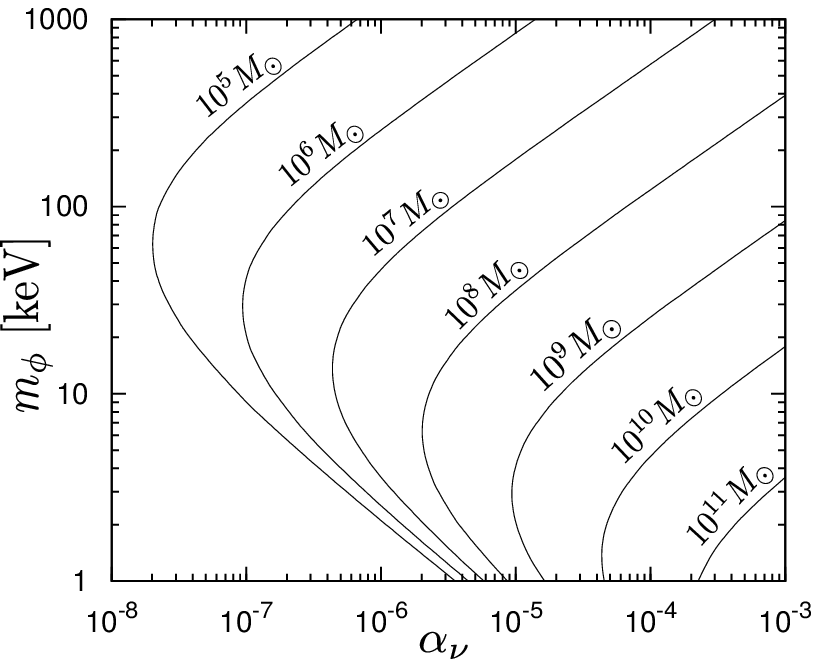}
        \caption{Exact numerical result}
        \label{fig:exactmassivepseudovector}
    \end{subfigure}
     \begin{subfigure}[htb]{0.49\textwidth}
        \includegraphics[scale=0.9]{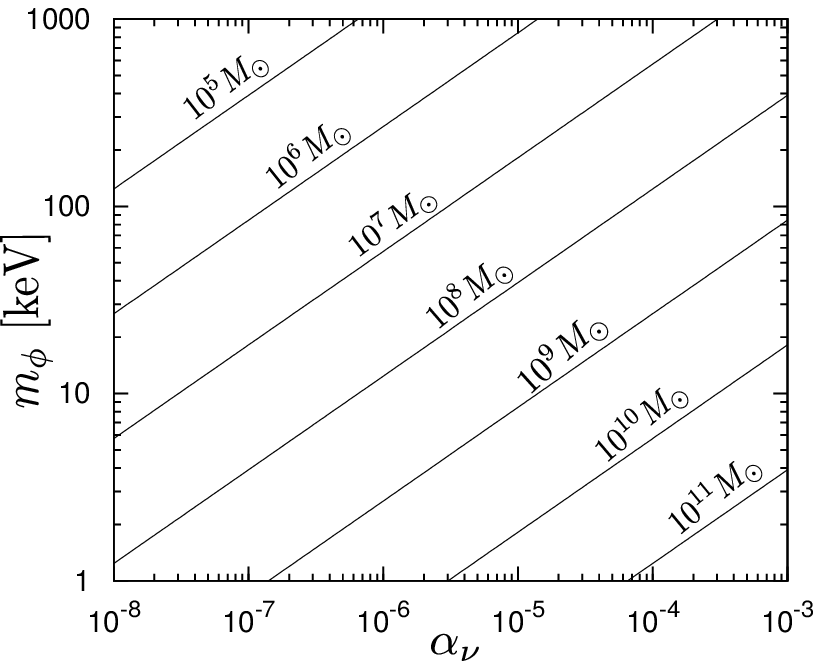}
        \caption{Effective propagator description}
        \label{fig:effectivepseudovector}
    \end{subfigure}
\caption{Contour line of a constant $M_{\text{cut}}$ is shown for the pseudo vector operator within the exact (left) and effective propagator framework (right). 
The DM parameters chosen are $m_{\chi}=100$\,MeV and $\alpha_{\chi}$ satisfying the relic density constraint. 
The effective description is valid only in the upper right quarter of figure~\ref{fig:effectivepseudovector}. }
\label{fig:massivepseudovectormcuts}
\end{figure}

Using the effective propagator description, we derive $\gamma/H$: 
\begin{align}
\frac{\gamma}{H}= 2.1 \times \left(\frac{r}{r_0}\right)^2 \left(\frac{N_{\nu}}{6}  \frac{\alpha_{\chi}}{8.4 \times 10^{-12}}\frac{\alpha_{\nu}}{10^{-4}}\right) \left( \frac{m_{\chi}}{100 \,\text{MeV}} \right)^{-1}\left( \frac{m_{\phi}}{100 \,\text{keV}} \right)^{-4} \left( \frac{T_{\nu}}{1 \,\text{keV}} \right)^{4} \,,\label{eq:gammahpv}
\end{align}
and the cutoff mass scaling pattern for pseudo vector interaction:
\begin{align}
(M_{\text{cut}})_{\text{PV}}= 1.4 \times 10^8  \, M_{\odot} \left(\frac{r}{r_0}\right)^{15/8} \left(\frac{N_{\nu}}{6} \frac{\alpha_{\nu}}{10^{-4}} \right)^{3/4} \left( \frac{m_{\chi}}{100 \,\text{MeV}} \right)^{-3/2}\left( \frac{m_{\phi}}{100 \,\text{keV}} \right)^{-3/2} \,, 
\label{eq:mcutpv}
\end{align}
where the relic density constraint on $\alpha_{\chi}$~(\ref{eq:relicabundancepv}) is inserted into (\ref{eq:gammahubbleexact}). 
Note that the cutoff mass depends now on the DM mass unlike in the scalar and vector operator cases. 
In figure~\ref{fig:massivepseudovectormcuts}, the exact numerical solution of $\gamma/H$ is compared to the cutoff derived from (\ref{eq:mcutpv}), showing the valid range of the parameter space for an effective propagator description.



\end{document}